\documentclass[pdflatex,sn-basic,iicol]{sn-jnl}
\usepackage{url}

\usepackage{supertabular}
\usepackage{ragged2e} 

\RequirePackage{color}
\newcommand\apjs{ApJS}               
\newcommand\aap{A\&A}                
\newcommand\ao{Appl. Opt.}           
\newcommand\apjl{ApJ}                
\begin{document}

\title[Ultraviolet Spectropolarimetry of B stars]{Ultraviolet Spectropolarimetry: on the origin of rapidly rotating B stars}

\author*[]{\fnm{C.E.~}\sur{Jones$^{1}$}}\email{cejones@uwo.ca}

\author[]{\fnm{J.~}\sur{Labadie${\text -}$Bartz$^{2}$}}

\author[]{\fnm{D.V.~}\sur{Cotton$^{3,4,5}$}}

\author[]{\fnm{Y.~}\sur{Naz\'e$^{6}$}}

\author[]{\fnm{G.J.~}\sur{Peters$^{7}$}}

\author[]{\fnm{D.J.~}\sur{Hillier$^{8}$}\newline}

\author[]{\fnm{C.~}\sur{Neiner$^{9}$}}

\author[]{\fnm{N.D.~}\sur{Richardson$^{10}$}}

\author[]{\fnm{J.L.~}\sur{Hoffman$^{11}$}}

\author[]{\fnm{A.C.~}\sur{Carciofi$^{12}$}}

\author[]{\fnm{J.P.~}\sur{Wisniewski$^{2}$}\newline}

\author[]{\fnm{K.G.~}\sur{Gayley$^{13}$}}

\author[]{\fnm{M.W.~}\sur{Suffak$^{1}$}}

\author[]{\fnm{R.~}\sur{Ignace$^{14}$}}

\author[]{\fnm{P.A.~}\sur{Scowen$^{15}$}}


\abstract{ UV spectroscopy and spectropolarimetry hold the key to understanding certain aspects of massive stars that are largely inaccessible (or exceptionally difficult) with optical or longer wavelength observations. As we demonstrate, this is especially true for the rapidly-rotating Be and Bn stars, owing to their high temperatures, geometric asymmetries, binary properties, evolutionary history, as well as mass ejection and disks (in the case of Be stars). UV spectropolarimetric observations are extremely sensitive to the photospheric consequences of rapid rotation (\textit{i.e.} oblateness, temperature, and surface gravity gradients), far beyond the reach of optical wavelengths. Our polarized radiative-transfer modelling predicts that with low-resolution UV spectropolarimetry covering 120 -- 300 nm, and with a reasonable SNR, the inclination angle of a rapid rotator can be determined to within 5 degrees, and the rotation rate to within 1\%. The origin of rapid rotation in Be/n stars can be explained by either single-star or binary evolution, but their relative importance is largely unknown. Some Be stars have hot sub-luminous (sdO) companions, which at an earlier phase transferred their envelope (and with it mass and angular momentum) to the present-day rapid rotator. Although sdO stars are small and relatively faint, their flux peaks in the UV making this the optimal observational wavelength regime. Through spectral modelling of a wide range of simulated Be/n+sdO configurations, we demonstrate that high-resolution high-signal-to-noise ratio UV spectroscopy can detect an sdO star even when $\sim$1,000 times fainter in the UV than its Be/n star companion. This degree of sensitivity is needed to more fully explore the parameter space of Be/n+sdO binaries, which so far has been limited to about a dozen systems with relatively luminous sdO stars. We suggest that a UV spectropolarimetric survey of Be/n stars is the next step forward in understanding this population. Such a dataset would, when combined with population synthesis models, allow for the determination of the relative importance of the possible evolutionary pathways traversed by these stars, which is also crucial for understanding their future evolution and fate. }


\maketitle

\section{Introduction} 
\label{sec:intro}


The proposed MIDEX mission, Polstar, will provide high-resolution UV spectroscopy and spectropolarimetry and offer a unique opportunity to study massive stars in the 122-320 nm wavelength range with unprecedented detail (see \citet{ScowenTC} in this volume). In what follows, our work is motivated by these proposed observations but we emphasize that our predictions and analysis can be applied to any UV spectropolarimetric observations obtained with the appropriate resolution, for example with the proposed ESA M-class mission Arago \citep{pertenais2017} or the Pollux instrument proposed on the LUVOIR NASA flagship project \citep{ferrari2019}.

First discovered by \citet{Secchi1866}, emission-line B-type stars are formally defined by having observed emission in the hydrogen Balmer lines at some time. However, there are many different types of objects which meet this phenomenological description such as the Herbig Ae/Be stars, B[e] supergiants, B stars with strong magnetospheres, interacting binaries, etc. \citep[see Table 1 in][]{Porter2003}. In what follows, one of the types of stars we consider are the ``classical Be stars'' (hereafter Be stars) which are non-supergiant B-type stars that have exhibited this type of emission at some time \citep{Collins1987} which is due to a nebulous disk of gas ejected from the star \citep{1931ApJ....73...94S}. As the definition implies, these stars are variable over
a range of timescales. They are distinguished by their rapid rotation which is thought to be less than critical \citep[see][and references therein]{Rivinius2013}. 

The gaseous circumstellar disks of Be stars are formed from material ejected from the central star which, due to viscosity, slowly flows outward. Such disks, called viscous decretion disks (VDDs), were originally proposed by \citet{Lee1991}.
The VDD model has been successful in reproducing observations over a range of wavelengths that were obtained with different observational techniques (see, for example, \citet{Haubois2014, Vieira2017,Rimulo2018,Klement2019,Ghoreyshi2021,Marr2021}). Mounting evidence suggests that the disks rotate in a near-Keplerian fashion \citep{Rivinius2013}. These disks are also geometrically thin as conclusively demonstrated in an interferometric and spectropolarimetric study of seven Be star disks by \citet{Quirrenbach1997}. Radiative processes within the disks result in numerous observational effects, including line emission, excess continuum emission especially at infrared and radio wavelengths, linear polarization, and UV features from wind-disk interaction. All are generally variable in time, as the density and temperature profiles of the disk can evolve on timescales from hours to decades. The physical extent of these disks is ambiguous, since different observables probe different regions of the circumstellar environment. For instance, linear polarization and optical continuum emission are generally formed in the innermost parts of the disk (within a few stellar radii or less), while radio emission can arise from disk material out to several hundred stellar radii. Figure~12 in \citet{Haubois2012} shows the radial distances where the bulk of these observables are produced for an early Be disk model.

The exact mechanism(s) that result in the formation of the disk has been debated for decades
\citep[\textit{e.g.},][]{1931ApJ....73...94S}. There is no doubt that rapid stellar rotation, by reducing the effective gravity in the equatorial region, is a key ingredient. However, it is generally believed that these stars are not rotating critically, so at least one another mechanism must be acting to help launch material into orbit.
This could be pulsations excited stochastically in the stellar core. They may have a high amplitude due to the rapid rotation of the star and can transport angular momentum to the surface layers (for example, see \citealt{Baade2016,Baade2018,Neiner2020} and references therein). 
Magnetic fields have been suggested as a possible mechanism to facilitate the release of material to form a disk \citep[see][]{Balona2020}. However, there are problems with this proposition as no large-scale field has been found for any Be star (see \citealt{Wade2016} for efforts of the MiMeS survey). Moreover, \citet{UdDoula2018} showed that for a polar field strength above a few tens of Gauss, the magnetic field would disrupt the Keplerian disk, and so strong dipolar magnetic fields and Be disks are incompatible.

In this work, we also consider the Bn stars which have similar stellar properties to the Be stars (\textit{i.e.} rapid rotation and a wide range of spectral sub-types, usually B5 to A2), but have never been observed to have a disk. The `n' in `Bn' indicates nebulous absorption lines (usually metals), which are widened due to high $v\sin{i}$ \citep{Cochetti2020}. Despite the name, the origin of these lines is now understood as being purely photospheric.

Just like the Be stars, it is unknown how Bn stars acquired their present-day rapid rotation. Whatever physical mechanisms act in Be stars to create outflowing disks they must not operate (efficiently) in Bn stars (otherwise they would create disks and be known as Be stars). Alternatively, Bn stars might either exhibit emission sufficiently rarely (they could
simply be dormant Be stars) or possess only quite tenuous disks, so that they have not been classified as Be stars. Regardless, studying the population of Bn stars is important for understanding the rotational evolution of massive stars, and for elucidating the Be phenomenon -- the actual physical mechanism that ejects mass and angular momentum, and that is either lacking or operating at low efficiencies in Bn stars.

Stellar rotation is important during main sequence (MS) evolution since it influences mass loss and the mixing of chemical elements and may impact post-MS stages, although this remains unclear. Be stars may have achieved their rapid rotation from birth on the zero age main sequence (ZAMS) \citep{Martayan2006}, through single-star evolutionary processes \citep{Georgy2013}, via binary interactions \citep{Pols1991}, or by some combination of these.
The VDDs of Be stars transport angular momentum and mass out, and are important for understanding the current and future evolution of these objects. With time-series observations, the angular momentum and mass flux can be calculated, providing a unique probe of these evolutionary processes \citep{Rimulo2018,Vieira2017} which can then be used to calibrate stellar evolution models \citep[\textit{e.g.,}][]{Granada2013}.
  
The fraction of Be stars in binary systems is a matter of debate and a key aspect of this work (and other ongoing research). It has been suggested that all Be stars are in binary systems \citep{Kriz1975}. Yet, a study by \citet{Oudmaijer2010} found that only $\sim$30$\%$ of Be stars are found in wide binary systems (with orbital separations of 20–1000\,au), which is essentially the same estimate found for B stars. More recently, from a study of 340 Be stars, \citet{Miroshnichenko2016} concluded that the binary fraction is greater than 50\%. Furthermore, it is unclear what fraction of Be stars have close binary companions\footnote{``Close'' here is taken to mean that binary interaction has or will have occurred at some point over the main sequence (MS) or post-MS evolution of one or both stars.}. Over the past decades, it has been suggested that binary interaction causes the spin-up of the Be star when an evolved companion donates material \citep{vanBever1997,Shao2014}. \citet{Hastings2021} concluded that it is perhaps plausible for all Be stars to have been spun up by binary interaction. However, it is unclear if this is realized in nature given the somewhat extreme assumptions required (``an initial binary fraction very close to unity, a shallow initial mass function and very non-conservative mass transfer''). 

In a different approach towards the same question,  \cite{Bodensteiner2020} investigated the literature for 287 early-type Be stars and concluded that none of them have known MS companions, in contrast to a large fraction of non-rapidly rotating B stars that do have MS companions. This finding supports the idea that the binary channel is responsible for a majority of the rapidly rotating B stars.

Even when a companion is not directly detected, its gravitational influence on the outer parts of the disk (causing \textit{e.g.}, disk density changes or truncation) can be measured.
Analyzing the spectral energy distributions (SEDs) of 57 Be stars out to radio wavelengths, \citet{Klement2019} found evidence for disk truncation in all of the systems with sufficient data, in support of the hypothesis that a large majority of Be stars have been spun up by a companion. 
An additional method for indirectly inferring a companion is through disk oscillation modes in resonance with a binary orbit (\textit{i.e.} $m$=2 modes). These $m$=2 modes are predicted by models \citep{Panoglou2018,Cyr2020}, and manifest in optical spectroscopy as periodic variations in emission line profile shape with orbital period \citep[\textit{e.g.,}][]{Peters2008,Peters2016, Chojnowski2018}. Unfortunately, no comprehensive survey has been conducted to quantify how common these $m$=2 modes are, an effort that is complicated by the general variability of the disks (often preventing any sort of ``steady state'' from being realized). 

Finally, there are at most a few tens of Be stars where their binary status and their evolutionary journey to rapid rotation are fairly well known. These are the Be + subdwarf O star (sdO) binaries. In these binaries, the sdO star is the remnant core of an evolved star that has donated mass and angular momentum to the present-day Be primary, thus spinning it  up. Such binaries are notoriously difficult to detect and classify, as the contribution in the optical flux from an sdO binary companion typically amounts to (perhaps significantly) less than a few percent owing to the much larger physical size of the Be star \citep{Mourard2015}. 

\citet{Wang2018} examined IUE spectra of 264 Be stars and found 12 new candidate  Be+sdO  systems (using cross-correlation techniques) of which 9 were later confirmed using HST observations \citep{Wang2021}. As the known Be+sdO systems have spectral classes B0-B3, 
\cite{Wang2018} argue that there must be many Be+sdO systems that cannot be detected using existing observations and current techniques.

In contrast, there are also works that suggest the validity of single-star evolutionary channels to rapid rotation \citep{Ekstrom2008,Granada2013}, without the need to invoke binary evolution. \cite{vanBever1997} determined that only $\sim$5 -- 20\% of Be stars can be formed through binary interactions, while \citet{Pols1991} determined that about half of the Be star population is the result of binary evolution.

Analysis of the B and Be star populations of clusters seems incompatible with single-star evolution being the dominant channel for acquiring rapid rotation \citep{McSwain2005}.
Based on the observed fraction of runaway stars in a large sample of Be stars, \citet{Boubert2018} suggest that all Be stars could be products of binary mass transfer.
In any case, the fraction of Be/Bn stars that have acquired their rapid rotation via binary interaction versus single-star evolution remains an open question. 

Stellar evolution models for single stars have been developed that predict how rotation rate evolves over time, for a range of initial rotation rates \citep{Eggenberger2008, Georgy2013}. Consequently, accurately measuring the rotation rate -- along with other physical parameters such as mass and radius -- in both Be and Bn stars constitutes a direct test of the feasibility of rapid rotation from single-star formation scenarios.

We further detail the motivation for our investigation in Section~\ref{motivation} and discuss the required observations, including our target selection, in Section~\ref{observations}. Evolutionary  considerations and potential results are described in Sections~\ref{evol} and \ref{results} respectively. The specific objective for this work is to test the prediction that the spin-up of massive main-sequence stars to form rapidly rotating star plus disk systems is caused by binary mass transfer, followed by rotational evolution due to core contraction and mass loss. To accomplish this objective we subdivide our results as follows: details and techniques about detecting binary companions and their orbits are provided in Subsection~\ref{binarydetails}, the modeling of the sdO stars is presented in Subsection~\ref{sdO_models}, how we determine rotational properties of the sample is outlined in Subsection~\ref{rotation}, and disk physical properties are described in Subsection~\ref{disks}. We conclude in Section~\ref{conclude}. A provisional target list pertinent to the capabilities of the {\em Polstar} mission has been assembled and is provided in Appendix A. Appendix B explores the relationship between the UV spectroscopic  signal-to-noise ratio (SNR) and the detectability of sdO companions in various binary configurations.

\section{Motivation}
\label{motivation}

The purpose of this work is to demonstrate how UV polarization and spectroscopy can uniquely address 
fundamental questions regarding the most rapidly rotating massive stars. 
Understanding the origins of rapid rotation requires knowledge of the stellar (and binary) evolutionary history and the present-day rotation rates.

\begin{figure}
    \centering
    \includegraphics[width=0.49\textwidth]{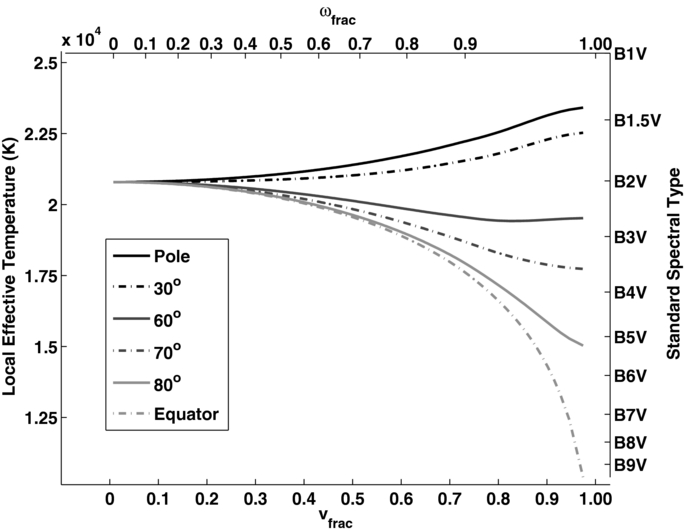}
    \caption{The stellar effective temperature with changes in stellar rotation as a function stellar latitude. The top axis gives the angular velocity as a fraction of the critical angular velocity and the bottom axis gives the velocity as a fraction of critical velocity, $v_{crit}$. The axis on the right gives spectral type corresponding to the surface temperature of a non-rotating B star. The rotating star is an early Be star of spectral subtype of B2. Reproduced with permission from \citet{McGill2011}, figure 1.}
    \label{gravitydark}
\end{figure}

\begin{figure}
    \centering
    \includegraphics[scale = 2.8]{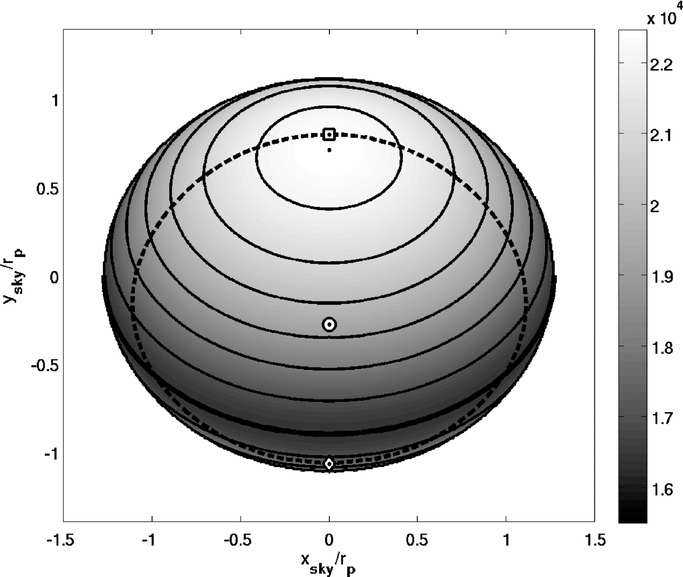}
    \caption{An early Be (B2) star rotating at 0.8 $v_{crit}$ with the corresponding surface temperature as a function of latitude given by the color code on the right hand side. The dashed line shows the portion of the stellar surface that could be viewed at radial distance of 2.0 stellar polar radii and at a height about the equatorial plane of 1.2 polar radii. Reproduced with permission from \citet{McGill2011}, figure 2.}
    \label{shape}
\end{figure}

Figures~\ref{gravitydark} and ~\ref{shape} demonstrate the importance of determining stellar rotation rates. Figure~\ref{gravitydark} shows the surface temperature variations with the rotation rate for an early B2 spectral type star. As critical rotation is reached, the temperature variation across the surface covers virtually the entire range for B stars. 

Figure~\ref{shape} shows the same spectral type (B2) star rotating at 0.8 $v_{crit}$ whose surface has an oblate shape due to the rapid rotation. It also illustrates that the projected area of the star and stellar surface temperature vary with viewing angle. Given the star is the main energy source for the disk
through its ionizing radiation, the requirement to obtain accurate rotation rates is important and has further implications for observables.

Polstar provides the capabilities for different (and generally complementary) observational strategies to determine certain properties of a large sample of Be/Bn stars and their circumstellar environments (discussed in Subsection.~\ref{disks}). Currently, the actual rotation rates are still contentious and the lack of firm rates means that we do not know actual stellar momenta. For example, \citet{Townsend2004} argue that gravity darkening due to the oblate stellar surface has led to a systematic underestimate of stellar rotation rates. These rates are particularly important to stellar evolutionary models and the evolution of the stars themselves. For a large sample, UV linear polarization with Polstar will provide rotation rates with extremely high precision{\bf ,} compared to conventional methods (\textit{e.g.,} optical spectral line fitting). 

UV data will allow the direct detection of companions. Indeed, the direct detection of hot sdO companions of Be/Bn stars, to date, relied almost exclusively on UV spectroscopy. This is due to a combination of factors, including the high $T_{\rm eff}$ (hence a flux peaking in the UV) and low intrinsic luminosity (owing to a small size) of sdO stars. The slow rotation rate of sdO stars,
compared to their Be companion, also facilitates their detection --  the rich narrow absorption line spectrum of the sdO star
can be more easily discerned. In the Be+sdO systems studied by \cite{Wang2021} all sdOs except one (which has $v\sin{i} = 102$\,km\,s$^{-1}$) have $v\sin{i} < 40$\,km\,s$^{-1}$, while the Be stars all have $v\sin{i} > 250$\,km\,s$^{-1}$. These differences in rotational velocities are a direct consequence of binary mass transfer.

Furthermore, spectropolarimetry will provide information about the geometry, velocity, and inclination angle of the Be star disks, which can be used to determine the orbital inclination{\bf ,} thus allowing masses to be determined. High quality UV spectroscopy, spectropolarimetry, and continuum polarimetry of a large sample of rapidly rotating B stars thus has the potential to answer the question of the origin of rapid rotation{\bf ,} by determining the fraction of such stars spun up by binary interaction versus those that have evolved as single stars.

During this investigation, we plan to determine the relative importance of the two main evolutionary channels (single-star versus binary interaction) by which hot stars acquire rapid rotation. In order to answer this question, we also need to
study the incidence rates and properties of stripped binary companions to Be
and Bn stars with UV spectroscopy. Precise stellar parameters of the rapidly rotating hot stars also need to be accurately determined, as well as the distortions in geometry, surface gravity, and effective temperature on the stellar surface caused by rapid rotation. The similarities, differences, and evolutionary links between rapidly rotating stars that form disks versus those that do not will also advance our knowledge.

\section{Sample selection and proposed observations}
\label{observations}

\subsection{The Be and Bn populations and sample selection}
\label{sec:Be_Bn_pop}

The Be phenomenon within our Galaxy is found in stars between spectral types of approximately O9 through A0, but peaks around B2. In the SMC, with its lower metallicity, earlier O stars also exhibit the Be phenomenon \citep{Golden-Marx2016,Li2018}.  In the Be stars, rotation rates over the entire population are on average $\gtrsim$80\% of critical \citep{Fremat2005}, but with considerable spread such that Be stars as a class cannot be considered as critical rotators. 
The population of Bn stars is generally found to be between B5 and A2, but a small number of early-type Bn stars exist.
The Bn stars roughly have the same rotation properties as Be stars when considered according to spectral sub-type \citep[\textit{e.g.,}][]{Cochetti2020}.

While the Be star population has in common rapid rotation and the presence of a disk (at some point in their observational history), there is considerable diversity in the stellar pulsational properties \citep{Labadie-Bartz2020} and the properties of the disks \citep{LabadieBartz2018}. The general trend among Be star disks includes a decrease in the disk mass and density towards later spectral types \citep{Vieira2017}. The disks of mid- and late-type Be stars are often stable on timescales of years or decades. In contrast, the disks of early-type Be stars are much more prone to variability as a consequence of (sometimes highly) variable stellar mass ejection rates. Despite these trends, there is considerable diversity in disk properties and variability among early-type Be stars. For example, the disk of $\gamma$ Cas (B0.5Ve) has been built up at a relatively steady rate over a few decades and as of 2021 is relatively massive \citep{Pollmann2021}. On the other hand, HD 49330 has the same spectral type as $\gamma$ Cas but is highly variable as the stellar mass ejection turns on and off over months/years \citep{Huat2009}.

Considering the diversity of Be/Bn stars, lessons learned from a particular spectral sub-type or a small sample can only offer an incomplete picture of these populations.
A more comprehensive understanding of the relevant physics (\textit{i.e.}, rapid rotation and ties to stellar evolution, angular momentum transport, mass ejection) requires systematic studies of sufficiently large samples.

\begin{figure}
    \centering
    \includegraphics[clip, trim={0.06cm, 0.05cm, 0cm, 0.cm}, width=8.0cm]{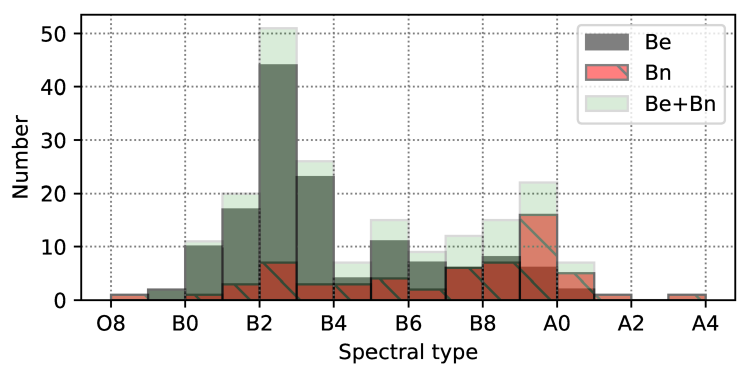}
    \caption{Distribution of the spectral types of the sample. Be and Bn stars are distinguished by different shading.}
    \label{fig:ST_hist}
\end{figure}

\begin{figure}
    \centering
    \includegraphics[clip, trim={0.06cm, 0.05cm, 0cm, 0.cm}, width=8.0cm]{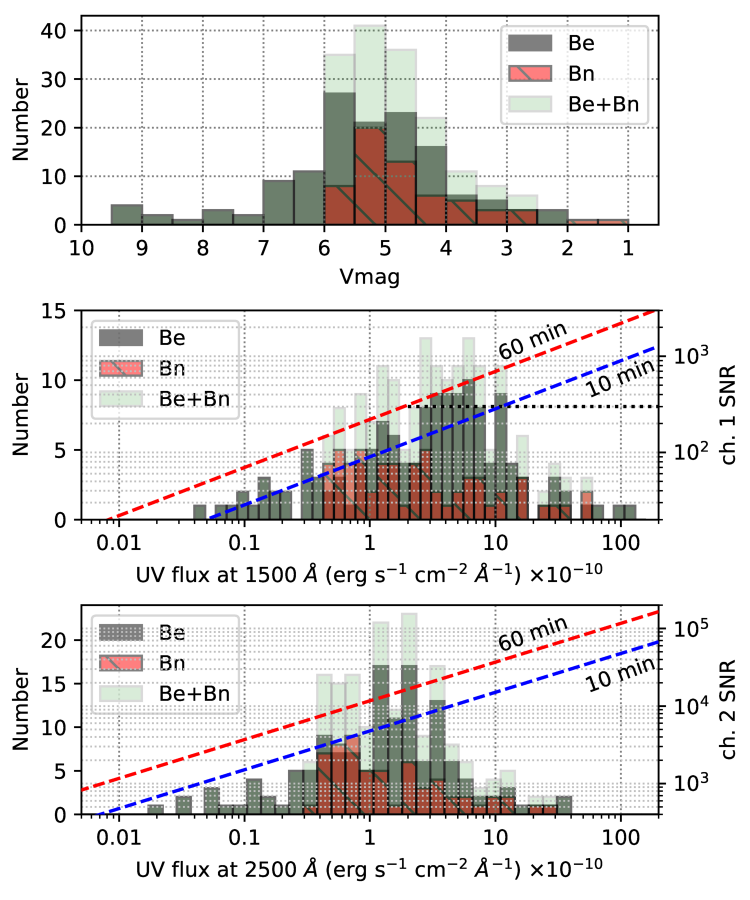}
    \caption{Histograms showing the brightness distribution of the target list, in terms of $V$-band magnitude (top), and UV flux at 1500 \AA  \,(middle) and 2500 \AA \,(bottom). In the two lower panels, the dashed lines show the expected Polstar SNR with 10 and 60 minute exposure times. }
    \label{fig:brightness_hist}
\end{figure}

Therefore, we have assembled a sample of 200 stars (140 Be, 60 Bn) that are representative of the general population. Histograms describing this sample are shown in Figure~\ref{fig:ST_hist} (spectral type) and Figure~\ref{fig:brightness_hist} (V magnitude and UV flux). Figure~\ref{fig:brightness_hist} also shows the expected SNR in Polstar's channel 1 and 2 (see Subsection~\ref{uvobservations} for details about these channels) over the range in UV flux for the sample for 10 minute and 60 minute exposures.

This sample includes 19 of the so-called $\gamma$ Cas analogues, which are a sub-class of Be stars \citep[\textit{e.g.,}][]{Naze2018}. These systems emit particularly hard and bright X-rays, but otherwise resemble perfectly typical Be stars. $\gamma$ Cas analogues have only been found among the early-type Be stars ($\leq$B3). One scenario to explain this peculiarity relies on the presence of evolved companions \citep{Murakami1986, Postnov2017, Langer2020}. Up to now, the presence of companions has been indirectly detected for at least eight cases through the motion of the primary, Be, star \citep[and references therein]{Naze2022}. While the data indicate low masses for the companions, their exact nature remains unclear and debated. Polstar can address this question via direct detection and characterization (or upper limits) of any UV-luminous companions. 

Also included in our sample are the 15 confirmed Be+sdO binaries (with direct detection of the sdO star in UV spectroscopy). 
These systems serve as a control group for sdO recovery and characterization. In the majority of these, only a few UV spectra of sufficient quality have been  obtained, and additional observations are thus needed to sample the orbital period. Tight constraints on the stellar and orbital properties of these systems are expected from our study.

\subsection{Proposed UV observations}
\label{uvobservations}

The Polstar mission is planned to have two observing modes, channel 1 and 2. In channel 1, UV spectroscopy is recorded at R$\approx$30,000 from 122 nm to 200 nm and channel 2, with R$\approx$100-1000 (higher at the FUV end), covers from 122 nm to 320 nm. A single observation in either channel always includes six sub-exposures that cycle through the waveplate positions, providing the information needed to extract the Stokes parameters (I, Q, U, V) and enabling spectropolarimetry. Further details are given in \citet{Scowen2021}.  Observations in both channel 1 and 2 can be leveraged to determine certain still-unknown properties of the Be and Bn populations. 

The high-resolution farther-UV channel 1 observations are ideal for detecting and characterizing sdO binary companions of Be/Bn stars. The past binary interactions that have resulted in the present-day Be+sdO systems are fundamental in the evolution of these systems, and are the main reason the mass-gainer is rapidly rotating. However, only 15 Be+sdO systems have been confirmed, most of which are poorly characterized, limiting the current understanding regarding the prevalence of this evolutionary channel among the population at large.

\begin{figure*}[ht!]
\centering
\includegraphics[width=0.99\textwidth]{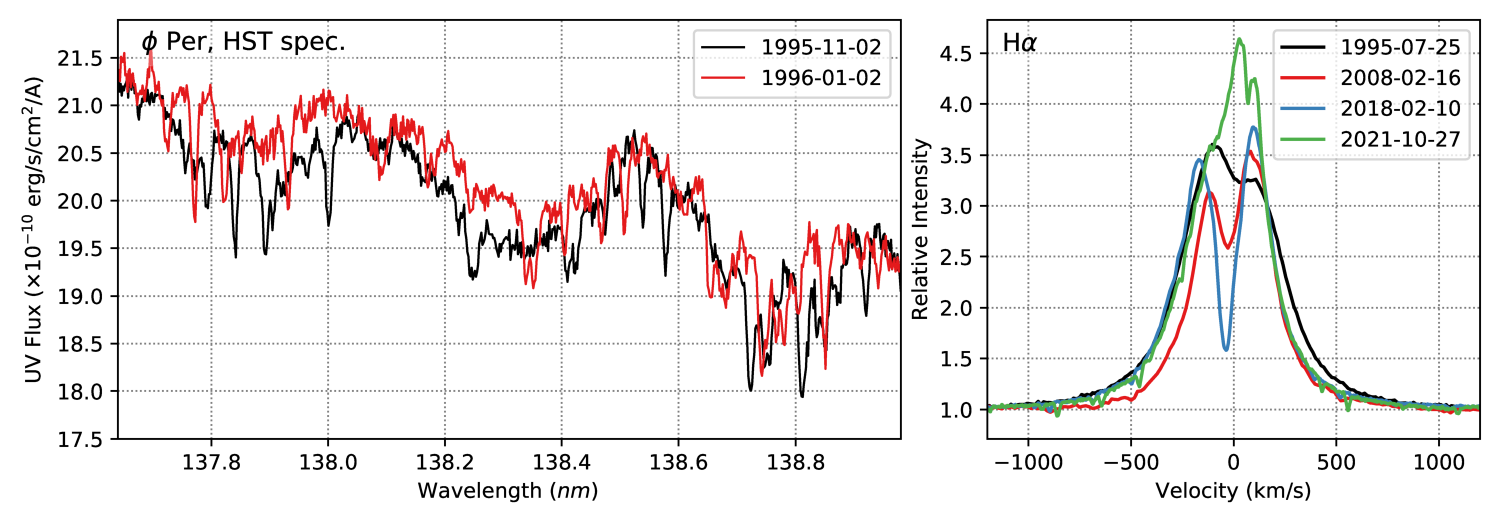}
\caption{
UV (left) and optical H$\alpha$ (right) spectra of the Be+sdO binary $\phi$ Per. The broad features in this region of the UV spectra are from the photosphere of the rapidly-rotating B star, and the narrow lines originate in the sdO photosphere (with clear RV variability at the two plotted epochs). The sdO stars are brightest in the UV, and the abundance of (relatively narrow) lines renders this spectral region optimal for sdO detection and characterization. The sdO star in $\phi$ Per is perhaps $\sim$5 -- 20$\times$ brighter than typical (being in a short-lived bloated stage), but has qualitatively the same spectral features as other sdO stars. The H$\alpha$ line traces a large volume of the disk (out to $\sim$10 $R_{\star}$), with the line profile containing information about the disk size, density profile, kinematics, inclination angle, and any asymmetries (in some cases induced by the sdO companion). The HST data were downloaded from MAST (\url{https://archive.stsci.edu/}), and the H$\alpha$ data from BeSS (\url{http://basebe.obspm.fr}).
}
\label{fig:phiPer}
\end{figure*}

Polstar (channel 1) is uniquely suited for conducting a UV spectroscopic survey of Be/Bn stars with sufficient precision, resolution, and cadence to detect hot sdO binary companions and determine the stellar properties and orbital parameters of the Be/Bn+sdO system. 
Towards this end, we propose $\sim$15 UV spectroscopic observations of nearly every star in the target list. Each observation should have SNR$\gtrsim$300 for at least the brightest $\sim$third of the sample, and then with SNR decreasing towards the fainter end of the distribution (to conserve the total exposure time). The need to detect even the faintest sdO stars (with sufficient signal to characterize their properties) drives the need for high SNR (see Section~\ref{sdO_models} and Figure~\ref{fig:BsdO_synspec}). 
The number of observations is motivated by the need to sample the orbital period (\textit{i.e.} to measure the orbital RVs of the sdO and Be/Bn stars). The cadence should be staggered so that orbital periods between $\sim$1 -- 10 months can be sampled. Most of the targets are not known to be binaries, but those that are have periods between $\sim$30 -- 200 days (excluding widely separated binaries which are not relevant in the context of binarity in stellar evolution).  

Also, for each target a single long exposure should be taken in channel 2 of Polstar. The target polarization precision in this observing mode is $\sim$2$\times 10^{-4}$.  Polarization measurements of the continuum in this wavelength region can provide the necessary information to determine rotation rates, inclination, and (latitude-dependent) surface temperature and gravity (see Section~\ref{rotation} for further details). Unlike the aforementioned goal, time-series information is not needed. 
Since the Bn stars do not have disks, their channel 2 polarization signal will be purely photospheric. The Be stars may or may not have a disk during their channel 2 Polstar visit. However, since the polarimetric signatures of a disk and an oblate star have different features, it is expected that these components can be separated (see Sec.~\ref{rotation} and~\ref{disks}).

To complement the UV data, we will organize a specific watch of our targets in the context of existing monitoring networks. For example, the Be Star Spectra (BeSS) observing network and database (initiated and run by one of us, C.N.) involves dozens of active observers who specialize in Be star spectroscopy and regularly contribute to peer-reviewed scientific publications \citep[\textit{e.g.,}][]{Naze2019, Richardson2021}. 
This combination of professional and amateur facilities will allow us to get optical spectra preceding and during the Polstar mission. The flexibility of this network will allow for time-series optical spectra to coincide with the Polstar observations of our targets. This strategy is currently being employed to monitor Be stars simultaneous with, \textit{e.g.}, the TESS space photometry mission \citep{Ricker2015}, and is easily adaptable to Polstar. Additionally, the ground-based observing network can be used to trigger UV observations in the event of a sudden mass ejection episode. These events are inherently unpredictable, but at the earliest stages imprint certain characteristic signals in optical data, and can thus (with a short time lag) be captured with Polstar. 

The Polstar data will reveal conditions on the stellar surface and near the central star, and any wind (and potentially any interaction of disk material with a companion). On the other hand, the visible spectra will allow us to consider the physical conditions in a larger volume of the gas, typically out to about 10 stellar radii.  This will allow us to determine the global properties of the disk (or lack thereof) at a given Polstar epoch, especially since the disks are variable in general and may not exist at all at certain times.

\section{Evolutionary Considerations}
\label{evol}
The origin of the Be phenomenon is still the subject of much debate. Is binarity essential for the
Be phenomenon? If not, then what fraction of Be stars are linked to binarity? Understanding the origin of Be stars is of course crucial for understanding their subsequent evolution.

Several studies have investigated whether Be stars can arise through single star evolution. A key process that can help facilitate the evolution of a young B star into a Be star arises because the maximum angular momentum a star can have (assuming rigid body rotation) decreases as the star evolves off the main sequence -- a consequence of the shrinking core \citep{Zhao2020}. As a result, the rotation rate of the evolving star may approach critical rotation when the material at the equator becomes unbound. Whether this actually occurs will depend on the efficiency of angular momentum transport and mass loss processes. Both of these are explicitly coupled to the rotation of the star. Two recent studies \citep{Hastings2020,Zhao2020} have both concluded that, while Be stars can arise from single star evolution, mass transfer in binary systems is important for Be star production. While rotation near critical has often been taken to be associated with the Be phenomenon, statistical \citep[\textit{e.g.},][]{Cranmer2005} and interferometric studies indicate that this is not needed --- many Be stars rotate at only 60 to 70\% of breakup.

Some evolutionary insights into Be+sdO binary systems can be gleaned from the best studied system, $\phi$ Persei, that contains a Be star with a mass of $\sim 9.6 \,M_\odot$ and an sdO star with a mass of $\sim 1.2\,M_\odot$ \citep{Mourard2015}. According to evolutionary calculations by \cite{Schootemeijer2018}, the masses of the original stars were  $7.2\pm 0.4\,M_\odot$ for the stripped (sdO) star and $3.8 \pm 0.4\,M_\odot$ for the Be star. If not for mass transfer, both stars would have ended their lives as CO WDs. However, since the mass of the Be star is now $\sim 9.6 \,M_\odot$ which is greater than the critical value of $\sim 8\,M_\odot$, it will most likely end its life as a Type IIP core-collapse SNe. The sdO star is very luminous for its mass, and must be He shell burning -- a stage that lasts for less than 3\% of the subdwarf's lifetime \citep{Schootemeijer2018}.

In general, the evolution of a binary star system depends fundamentally on the primary mass, the mass ratio, the orbital parameters, and the initial metalicity of the stars. Unfortunately, evolutionary calculations need to include numerous free parameters to facilitate the treatment of uncertain physical processes. The parameters govern (or set), for example, the efficiency of mass transfer, mass and angular momentum loss, the distribution of the kick velocity of any newly created neutron stars, the minimum mass for a core collapse supernova, etc. The study of binary systems can provide crucial constraints on some of the parameters. For $\phi$ Persei, \cite{Schootemeijer2018} found that the system must have evolved with nearly-conservative mass transfer. Despite the complexities of binary evolution, crucial insights into their evolution can be obtained from theoretical calculations.

In order to cover the large parameter space encompassed by binaries, and to test adopted parameters, an enormous number of models must be computed. For example, \cite{Zapartas2017} ran 3 million binary evolution calculations to investigate the late-time delay-time distribution of core-collapse SNe. Late-time SNe are those SNe that occur (in a coeval system) after all single stars more massive than $\sim 8\,M_\odot$ have exploded. Such SNe can occur up to 200~Myrs after the last single star explosion, and their existence requires binary star systems in which mass transfer has occurred. The \cite{Zapartas2017} study also provides insights into the evolution of Be+sdO systems.

In Be+sdO binaries the initial evolution is, of course, governed by the transfer of most of the H envelope of the (initially) more massive primary star to the secondary star.  The secondary now becomes the
more massive Be star. Eventually the secondary star will expand, and fill its Roche lobe, possibly triggering another phase of mass transfer and common-envelope evolution. Many different scenarios for the future evolution are possible -- some are outlined below.

(1) Merger of the He subdwarf with the evolved secondary star after a common envelope phase,
leading to a core-collapse SN. The merger process depends critically on the treatment of the common-envelope phase  \citep{Zapartas2017}. When the common envelope is ejected, the stars are less likely to merge. Significant progress towards 3D modeling of the common envelope phase is being made \citep[\textit{e.g.},][]{Lau2021}.

(2) Merger of the CO WD with the evolved secondary. The subsequent evolution is mass dependent, and uncertain. According to \cite{Zapartas2017} this could lead to a ONeMg core that may eventually collapse as a result of electron capture. Alternatively, the WD core could undergo a thermonuclear explosion inside a H-rich envelope. This could produce a Type Ia-CMS SN (\textit{e.g.}, SN2002ic, \cite{Hamuy2003}).  A final possibility is that the core never obtains sufficient mass to undergo core collapse.

(3) The sdO star can evolve into a CO (or potentially a ONe) WD. The secondary star can explode as a Type II SN, leaving a neutron star remnant. In most cases the neutron
star will be ejected from the system, however in some cases a WD + NS binary
system will be formed.

(4) The secondary star, even after it has accreted matter, may still be below the
threshold for a type II SN. In this case we end up with a binary WD system. Such
systems could potentially give rise to a Ia SN.

Until recently only a few systems like $\phi$ Persei were known. With the discovery
by \cite{Wang2018,Wang2021} of new Be+sdO systems, observations can begin to provide 
crucial constraints on binary evolution. Using Polstar we will be able to obtain
accurate orbital parameters and hence accurate masses for the confirmed Be+sdO systems, and for the many new systems we will find using the UV spectroscopic and polarization capabilities of Polstar.

\section{The Science Impact of {\em Polstar}}
\label{results}

\subsection{Determine the stellar and binary properties of the Be/Bn stars}
\label{binarydetails}

\subsubsection{The known Be+sdO population}
\label{Be+sd0}
High-resolution high-SNR UV spectroscopy is perhaps the best observational tool for the detection and characterization of hot small sdO companions to Be/Bn stars. To date there are 15 Be+sdO binaries where the sdO star has been directly detected in the UV \citep{Gies1998,Peters2008,Peters2013,Peters2016,Wang2017,Wang2018,Wang2021}.

The flux ratio at a given wavelength between an sdO and a Be star in a binary ($f_{\rm sdO} / f_{\rm Be}$) is a convenient parameter to determine the degree to which it is possible to de-convolve the composite spectrum as a function of SNR. Other stellar parameters ($T_{\rm eff}$, $\log{g}$, and $v\sin{i}$), plus binary motion, also have an effect. 

The Be+sdO system with the highest flux ratio, $f_{\rm sdO} / f_{\rm Be}$ $\approx$15\% (at $\sim$150 nm), is $\phi$ Per \citep{Gies1998}, where HST UV spectra revealed individual spectral features that are clearly seen to vary in RV (Figure~\ref{fig:phiPer}). However, the majority of the known Be+sdO binaries have a lower $f_{\rm sdO} / f_{\rm Be}$ of approximately 2.5\% -- 10\% \citep{Wang2021}. 

Other systems are known binaries (\textit{e.g.,} through SB1 orbital solutions for the Be star) but where no companion has been directly detected. For instance, an upper limit of $f_{\rm sdO} / f_{\rm Be}$ for the 203.5 d binary $\gamma$ Cas \citep[B0.5IVe,][]{Nemravova2012} is approximately 0.6\% at FUV wavelengths \citep{Wang2017}. Such binaries (with yet unseen companions) may also be found among the mid/late type B stars \citep[\textit{e.g.,} 7 Vul, B5Ve,][]{Harmanec2020}.  These and similar systems with known orbital periods are included in the target list, and Polstar observations can be scheduled to cover the orbital phases. There are additional numerous systems where RV motion of a Be star and/or its disk has been detected, but that so far lack an orbital solution \citep[\textit{e.g.,}][]{Chojnowski2017}. Thus, there are a large number of systems where binarity is suspected or known, but that are lacking in the observations needed to reveal the nature of the companion. 

There are very few, if any, confirmed Bn+sdO binaries. This is generally consistent with the lack of directly-detected sdO stars in binaries with mid- and late-type Be stars (since Bn stars are typically of later spectral types). However, Bn stars are often less enthusiastically observed compared to the Be stars, and the scarcity of known hot sub-luminous companions (either sdO/B or pre-WD) to Bn stars (and also later-type Be stars) is likely in part due to a lack of data sufficient for the task of detecting these faint sources. For example, Regulus, a Bn star (B8IVn), was recently found to host a pre-WD companion (\textit{i.e.} a low mass stripped core) in a 40 d orbit \citep{Gies2020}. 
The Regulus system likely has a similar binary evolution history as the Be+sdO binaries, yet has never been observed to build even a weak disk. With a flux ratio $f_{2}/f_{1} \approx 0.06\%$ in the visible, the UV flux from the pre-WD component of Regulus should be detectable in high-SNR channel 1 Polstar spectra.

\subsubsection{Expanding the Be/Bn+sdO parameter space by pushing to higher contrast systems and larger numbers}
\label{sec:BesdO_parameters}

The discovery of Be+sdO binaries has largely been driven by IUE UV spectroscopy \citep{Wang2018}. 
With these observations typically having low SNR (on the order of SNR$\sim$10), it is reasonable that the systems with relatively high values of $f_{\rm sdO} / f_{\rm Be}$ were the first to be found.
Although the known Be+sdO systems typically have $f_{\rm sdO} / f_{\rm Be}$ of a few percent, it is necessary to acquire observations that can probe flux ratios smaller than this. 
For example, with IUE spectra of 6 Be stars known to be binaries, \citet{Wang2017} derived upper limits on the flux contribution of any potential sdO companions to be $\lesssim$1\% for five stars, and detected an sdO star in one (60 Cyg, B1Ve). 
Thus, observations that 
are sensitive to flux ratios of $f_{\rm sdO} / f_{\rm Be} < 1 \%$ (and even down to  $\sim$0.1\%) are a necessary step towards a better understanding of this population. 

\begin{figure}
    \centering
    \includegraphics[clip, trim={0.06cm, 0.05cm, 0cm, 0.cm}, width=7.75cm]{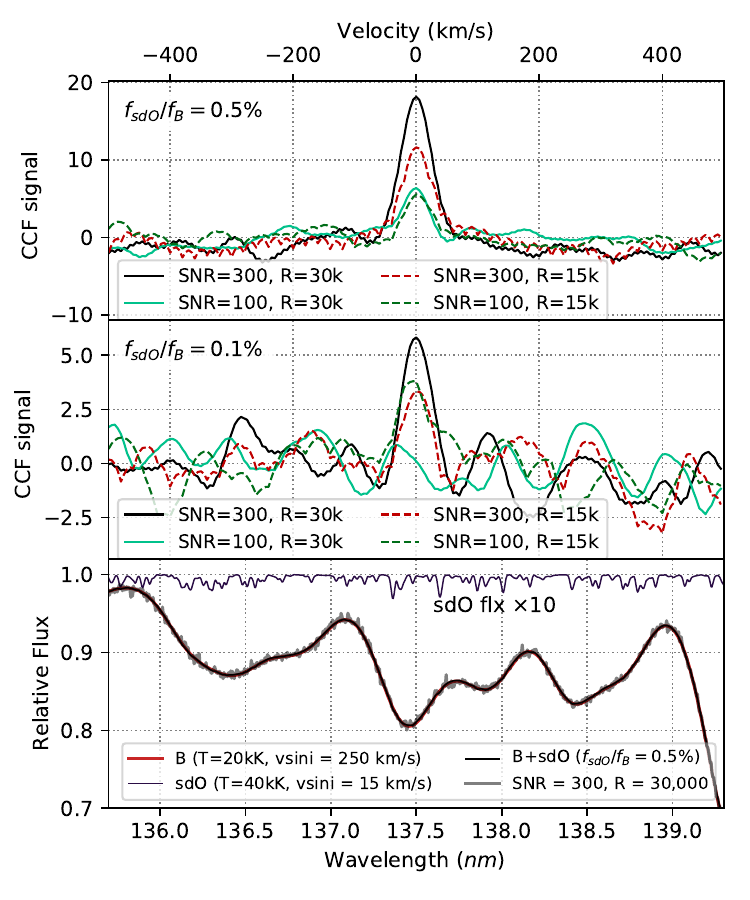}
    \caption{Cross-correlation function (CCF) signals for different combinations of SNR and resolution. The relative flux contribution from the sdO component ($f_{sdO}/f_{B}$) is 0.5\% in the top panel, and 0.1\% in the middle panel, and is centered at 0 velocity. The bottom plot shows a portion of the composite spectrum (for $f_{sdO}/f_{B}$ = 0.5\%), with the isolated sdO flux plotted above and scaled by a factor of 10. Typical rotation rates and temperatures are used for the B and sdO stars (as indicated in the bottom panel).  At such small contrast ratios, CCFs are needed to detect the sdO star, since individual lines are not easily identified in the composite spectrum.}
    \label{fig:BsdO_synspec}
\end{figure}

In a single Polstar channel 1 spectrum with R = 30,000 and SNR = 300, an sdO companion with a flux ratio as low as $f_{\rm sdO} / f_{\rm Be} \approx 0.1$\% may be detected (Figure~\ref{fig:BsdO_synspec}, middle panel). In systems with flux ratios of $f_{\rm sdO} / f_{\rm Be} \gtrsim 0.5$\%, a single observation of SNR$\sim$300 provides the means for a precise determination of the sdO properties (Figure~\ref{fig:BsdO_synspec}, top panel), and with multiple such observations the orbital properties can be determined. Then, with knowledge of the orbital RVs, each spectrum can be shifted to the sdO rest frame and co-added for perhaps a significant increase in the precision with which the sdO properties can be determined. 
Appendix~\ref{sec:detect_sdO} describes how UV spectroscopy, such as with Polstar, is capable of detecting faint sdO companions with a range of parameters.

Since $f_{\rm sdO} / f_{\rm Be}$ is generally small, cross-correlations functions (CCFs) or similar techniques to maximize signal typically by combining information from many lines need to be employed for sdO detection and characterization in Be+sdO binaries. CCF techniques were used to find candidate Be+sdO binaries from IUE spectra \citep{Wang2018}, and later to confirm many of these candidates with higher quality HST spectra \citep{Wang2021}. Details of this technique can be found in the aforementioned works, and the method is also illustrated in Figure~\ref{fig:BsdO_synspec}. In brief, grids of model sdO spectra (varying the RV and the stellar parameters) are cross-correlated to the observed spectrum (after correcting for the broad features of the rapidly rotating B star). The template that provides the best fit to the data then describes the stellar properties and velocity of the sdO star at each observed epoch.

To measure the orbital properties, multiple observations are needed to cover the orbital phase. In some cases, the orbital ephemeridies are already known even when a companion has never been directly detected (\textit{e.g.,} $\pi$ Aqr, $\gamma$ Cas). However, in the general case where no specific prior information about a binary orbit is known, observations spread out over weeks and months are likely to sample a binary orbit. In the known Be binaries (both with and without detected sdO companions), orbital periods range between a few weeks to a few months.

With Polstar, we aim to observe a sample of $\sim$140 Be and $\sim$60 Bn stars $\sim$15 times each with high-resolution (R = 30,000) UV spectroscopy. For the brightest $\sim$1/3rd of the sample, each observation should have SNR $\gtrsim$300 to enable precise measurements of the sdO properties and velocities at each epoch. In the remainder of the sample, a SNR of $\sim$50 -- 300 can be achieved with reasonable exposure times, which in most cases should be sufficient to detect sdO companions similar to those already known. In these relatively lower SNR spectra, any detected sdO stars can still be characterized by stacking multiple observations, provided that some information about the RV motion of the sdO star can be extracted.

With Polstar, we can obtain not only the number of these objects with compact sdO-type companions, but also the orbital inclinations with polarization data mainly from disk scattering.
In the majority of the known Be binaries the orbits are (very nearly) circular and probably co-planar with the rotation axis of the Be star (and thus also the Be star disk). When orbits are not aligned with the disk, tidal and radiative forces from the companion influence or warp the disk in characteristic ways, such that these misaligned cases can be additionally diagnosed through optical spectroscopy \citep[{\textit{e.g.,}}][]{Marr2021,Nemravova2010}, and eccentricity can be determined from the RV measurements (especially of the sdO star) as in \citet{Peters2013}.
When combining the polarization-derived inclinations and the double-lined orbits, we obtain a distribution of the masses for both the Be/Bn stars and the companion sdO/(pre-)WD stars. This will lead to powerful constraints for evolutionary pathways.

The combinations of channel 1  and channel 2 for our proposed targets are complementary for diagnosing the stellar properties of the Be/Bn stars. For instance, disentangling the B star rotation axis and its rotational velocity is enabled, since channel 1 spectroscopy and channel 2 spectro-polarimetry both constrain $v\sin{i}$ in different ways (see Sec.~\ref{rotation}).

\subsection{Model the spectra of sdO stars}
\label{sdO_models}

Due to the high quality spectra obtained with Polstar, and multiple observations, we will be able to extract the spectra for many of the sdO companions. This will, in turn, facilitate a direct analysis of the companion star using non-LTE radiative codes such as TLUSTY \citep{Hubeny1995} and CMFGEN \citep{Hillier1998}. Because of the narrow lines in sdO stars, we should be able to obtain accurate spectral parameters and abundances, which aid in the understanding of the past and future evolution of the system. Assuming observations with a signal-to-noise ratio of 1000, a sdO star with a flux of 10\% of the Be star will have (ignoring errors arising from disentangling the spectra) a signal-to-noise ratio of 100 (and \textit{e.g.}, for a relative flux of 1\% the SNR is 10).

\citet{Wang2021} used the height of the peak in the cross-correlation function to constrain the effective temperature of the sdO stars. They assumed a fixed gravity, { \bf $\log{g}$} of 4.75, and estimated  effective temperatures generally between 38000 K and 45000K (one had $T_{\rm eff}=33800$\,K) with an estimated error of 2500\,K. For the cross-correlation analyses they used theoretical spectra computed with TLUSTY. One issue with the models (and O and B stars in general) is that UV line lists are incomplete/inaccurate, and hence lines are missing (if we use only lists with accurate wavelengths) or at the wrong location (if we also include lines with theoretical wavelengths). Our understanding of the UV should improve over the next few years through the ULLYSES project and subsequent analyses. The ULLYSES Project (\url{https://ullyses.stsci.edu}) devoted 1,000 HST orbits to obtain a high quality UV spectral library of high and low mass stars.

There are several different methods available to extract individual spectra of the component stars in a binary system \citep[\textit{e.g.},][]{Simon1994,Gonzalez2006,Binnenfeld2020,Quintero2020}, although these have generally been designed to work in the optical, and they make assumptions not necessarily applicable to UV analyses. For example, the algorithm of \citet{Simon1994} works with rectified data, however the severe line blanketing in the UV makes it extremely difficult to rectify spectra. Some procedures suffer by convergence issues, although these can be addressed through modifications of the algorithm \citep[\textit{e.g.},][]{Quintero2020}. It is unclear however, if these procedures will work in the UV where the Fe forest causes severe line blending. Consequently we will need to test these algorithms, and as needed, develop modifications to improve their applicability to the UV, and to cases where the flux of the companion star is very much less than that of the Be star.

\subsection{Model the rotational properties of the sample}
\label{rotation}

Channel 1 spectroscopy will provide $v\sin{i}$ measurements for rapidly rotating stars (where $v$ is surface rotational velocity and $i$ is inclination of the stellar rotation axis to the line of sight). Combining Channel 2 continuum polarization will enable the inclination degeneracy to be overcome and the rotation rates of up to $\sim$200 stars to be determined (140 Be, 60 Bn). The modelling process to determine rotation rate and inclination also provides determinations of effective temperature, $T_{\rm eff}$, surface gravity, $\log{g}$, and subsequently mass, $M$ and radius, $R$.

As the rotation rate of a star increases it becomes increasingly oblate. Under a Roche model, at critical rotation, the equatorial radius is 1.5 times larger than the polar radius \citep{Meader2015} -- a value slightly modified by radiation pressure \citep{Gagnier2019}. Furthermore, as discussed above and shown in Figures~\ref{gravitydark} and~\ref{shape}, the temperature of the star varies with latitude, with the poles being hotter than the equatorial regions, which are said to be gravity darkened \citep{Meader2015, Gagnier2019}.

Along with other fundamental parameters like temperature and luminosity, the rotation rate, $\omega / \omega_c$ -- where $\omega$ is the angular rotational velocity, and the $\omega_c$ the critical angular velocity for break-up, probes a star's evolutionary status. In particular, a comparison is facilitated with the rapidly-rotating stellar evolution models of \citet{Eggenberger2008}, and \citet{Georgy2013}. These are single star models that simulate evolutionary tracks based on assumed initial mass and rotation rates, where deviation from the observations implies a binary merger or other binary interaction.

Ordinarily, without prominent surface features, it is not possible to determine the rotational velocity of a rapidly spinning B- or A-type star. Spectroscopy can only reveal the projected velocity, from its nebulous spectral lines, in the form of $v\sin{i}$. Two methods exist for overcoming this restriction. With the addition of readily available astrometric and spectroscopic information, both facilitate comparison to evolutionary models. 

The first method involves using interferometry to directly resolve the rotationally distorted shape of the stellar disk. Such an approach is intrinsically limited to the nearest stars. To date, several stars have had their rotational velocities measured in this way, and only four of them have spectral types of A5 or earlier: Achernar (a Be star) \citep{DdSouza2014}, Regulus (a Bn star) \citep{Che2011}, 51~Oph \citep{Jamialahmadi2015} and Rasalhague \citep{Zhao2009} -- three of which are closer than 50\,pc.

In the past five years a second approach using polarimetry has been demonstrated \citep{Cotton2017, Bailey2020}. Polarization from a hot stellar photosphere arises primarily as a result of electron scattering; it increases from the centre of the disc to the limb, to which its angle is tangential \citep{Chandrasekhar1946}. All of the polarization vectors cancel out for a spherical star, but for a rapid rotator a net polarization is produced as a consequence of the different environments at the poles and equatorial regions \citep{Harrington1968}.

The local polarization level depends on the ratio of scattering to absorption, and thus is larger for higher temperatures, and lower surface gravity. The net polarization further depends on rotation rate and the inclination to the line of sight. At visual wavelengths this polarization is small \citep{Collins1970, Sonneborn1982, Cotton2017, Bailey2020}. Within 100~pc of the Sun, the interstellar polarization level is similar to that induced by rotation in a B- or A-type dwarf star, but beyond, it is much larger and obfuscates the stellar intrinsic polarization.

\begin{figure}[ht]
    \centering
    \includegraphics[clip, trim={0.6cm, 0.5cm, 0cm, 1.5cm}, width=7.75cm]{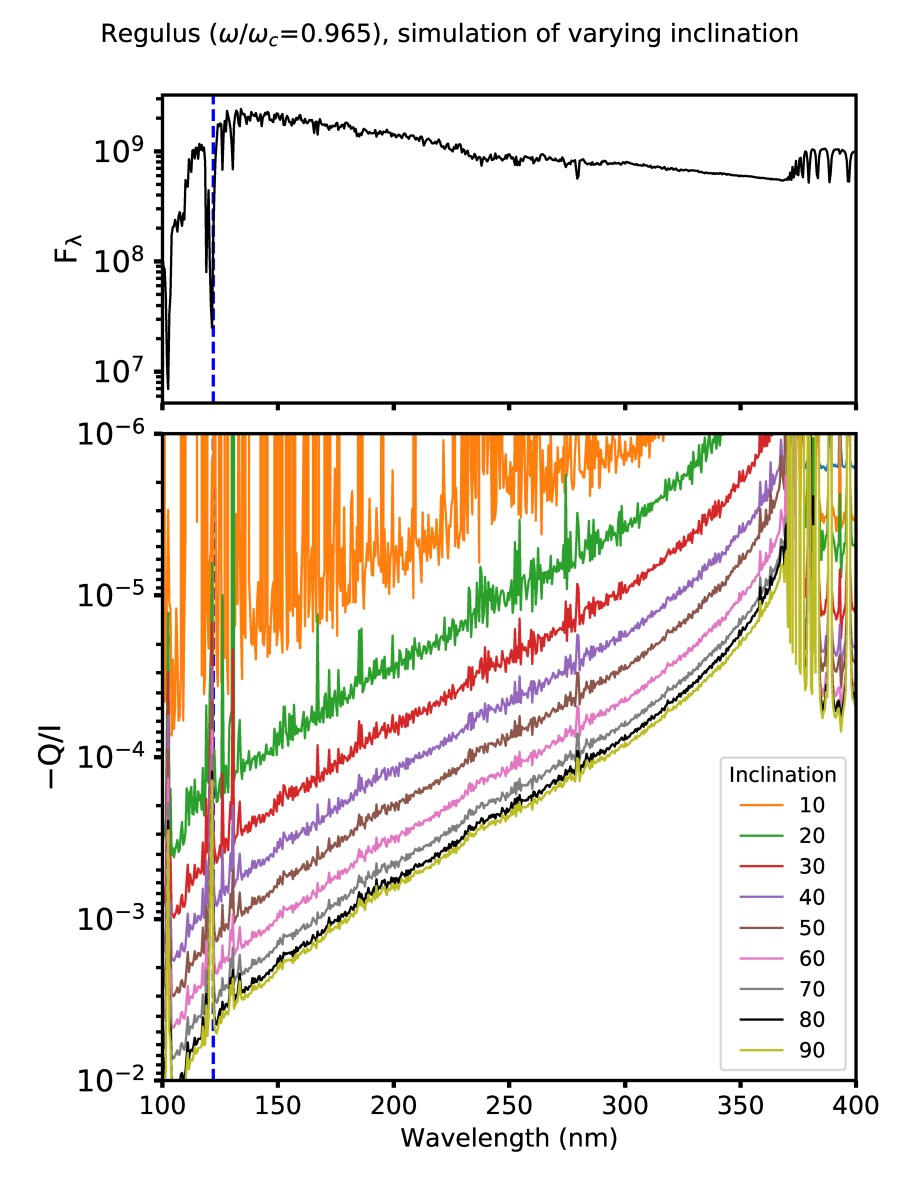}
    \caption{Simulations of differing inclination for the polarization of Regulus. The modelled surface flux, $F_\lambda$, in units of erg cm$^{-2}$  s$^{-1}$\AA$^{-1}$ (a) and polarization (b) UV spectrum of Regulus. The polarization is given in fractional units on a log scale with the most negative $Q/I$ values ({\textit i.e.} greatest polarization) at the bottom; the optical polarization of Regulus is at the level of a few $\times10^{-5}$  \citep{Cotton2017}. Regulus has $\omega/\omega_c=96.5\%$ and $i=80^{\circ}$ (black); other inclinations are simulated (colours). The polarization is presented in terms of normalized Q Stokes in the reference frame of the stellar rotation axis. In the UV the values are almost entirely negative in $Q/I$ -- perpendicular to the rotation axis -- with the largest negative values furthest into the UV. There is a marked difference in polarization with inclination -- with $i\ge20^{\circ}$ producing polarizations greater than $10^{-4}$ -- but in all cases the absolute value of polarization increases approximately logarithmically with decreasing UV wavelength after the Balmer jump. The dashed blue vertical line represents 122~nm. The plot is based on a \textsc{synspec/vlidort} model using the parameters determined by \citet{Cotton2017}. The models included the wavelength range from 100 to 400~nm and used the atomic UV line lists `gfFUV' and `gfNUV' acquired from the \textsc{synspec} website and originally computed by \citet{Kurucz1999}.}
    \label{fig:reg_uv}
\end{figure}

The distance limitation imposed by interstellar polarization can be greatly alleviated by working in the UV. At UV wavelengths the net polarization is orders of magnitude larger than in the optical \citep{Sonneborn1982, Collins1991, Lewis2022}. Gravity darkening at the equator increasingly exacerbates the inequality in fluxes from the polar and equatorial regions shortward of the equatorial blackbody peak. The enhanced UV polarization is demonstrated for Regulus in Figure \ref{fig:reg_uv}. Here, we can see that polarizations of up to $5\times10^{-3}$ are predicted for Polstar's wavelength range and that this value is highly wavelength dependant. For more luminous stars, the polarization can be even higher -- up to $10^{-1}$ or more for a giant with a spectral type later than A-type \citep{Lewis2022}. Consequently Polstar will be able to exploit this phenomenon to determine rotational velocities for stars at much greater distances.

Four parameters contribute to the shape and magnitude of the observable polarization curve. Two of these are global parameters: rotation rate, $\omega/\omega_c$, and inclination, $i$. The other two -- temperature and surface gravity -- vary over the stellar disc, but can be parameterized as their values at a single surface location, the pole for instance (as $T_p$ and $\log{g_p}$), if a gravity darkening law is assumed along with a Roche model for the stellar shape. The classical law is that of \citet{vonZeipel1924}, but the \textit{ELR Law} \citep{EspinosaLara2011}, which produces less gravity darkening, is preferred because it has been specifically designed to model rotating stars. With a known magnitude, distance and $v\sin{i}$, we can determine $T_p$ and $g_p$, if we assume $i$ and $\omega/\omega_c$. The methodology, then, is to construct a 2D parameter grid of assumed $i$ and $\omega/\omega_c$ values, where each point on the grid has a predicted polarization versus wavelength curve, which is then compared to observations \citep{Cotton2017, Bailey2020, Lewis2022}. In this way, all four parameters are determined -- $T_p$, $g_p$, $i$ and $\omega/\omega_c$. An inversion of \citet{Howarth2001}'s method for determining distance can then also be used to determine mass, $M$, and make comparison to evolutionary models. 

\subsubsection{\textit{Polstar}'s sensitivity to rotation induced polarization}
\label{sec:rot_calc}

\begin{figure}[ht]
    \centering
    \includegraphics[clip, trim={0cm, 0cm, 0cm, 0.cm}, width=7.5cm]{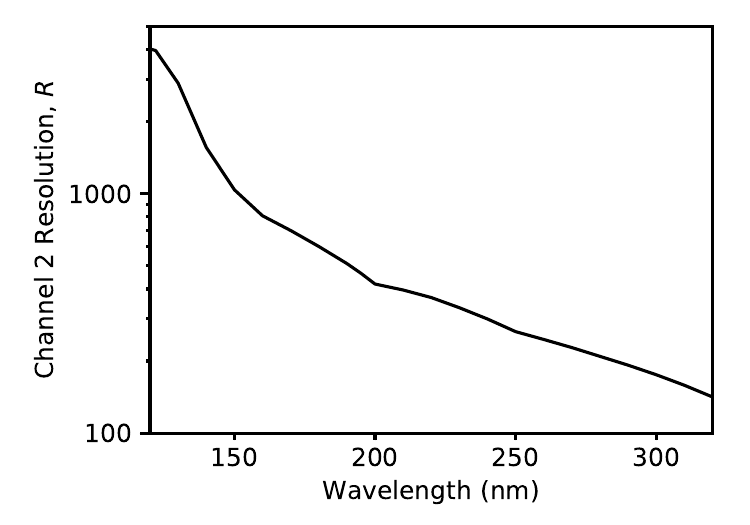}
    \caption{The spectral resolution of Polstar's Channel 2, as used for calculations in Section \ref{sec:rot_calc}.}
    \label{fig:Ch2R}
\end{figure}

Here we demonstrate Polstar's sensitivity to rotation-induced polarization by considering a Regulus-like star as an example. In the analysis that follows, spectra have been smoothed and binned corresponding to the current design for Polstar's Channel 2 -- the spectral resolution, $R$, is shown in Figure~\ref{fig:Ch2R}; the bin width is inversely proportional to 2.5$R$. 

\begin{figure*}[!tp]
    \centering
    \includegraphics[clip, trim={0cm, 0cm, 0cm, 0cm}, width=16cm]{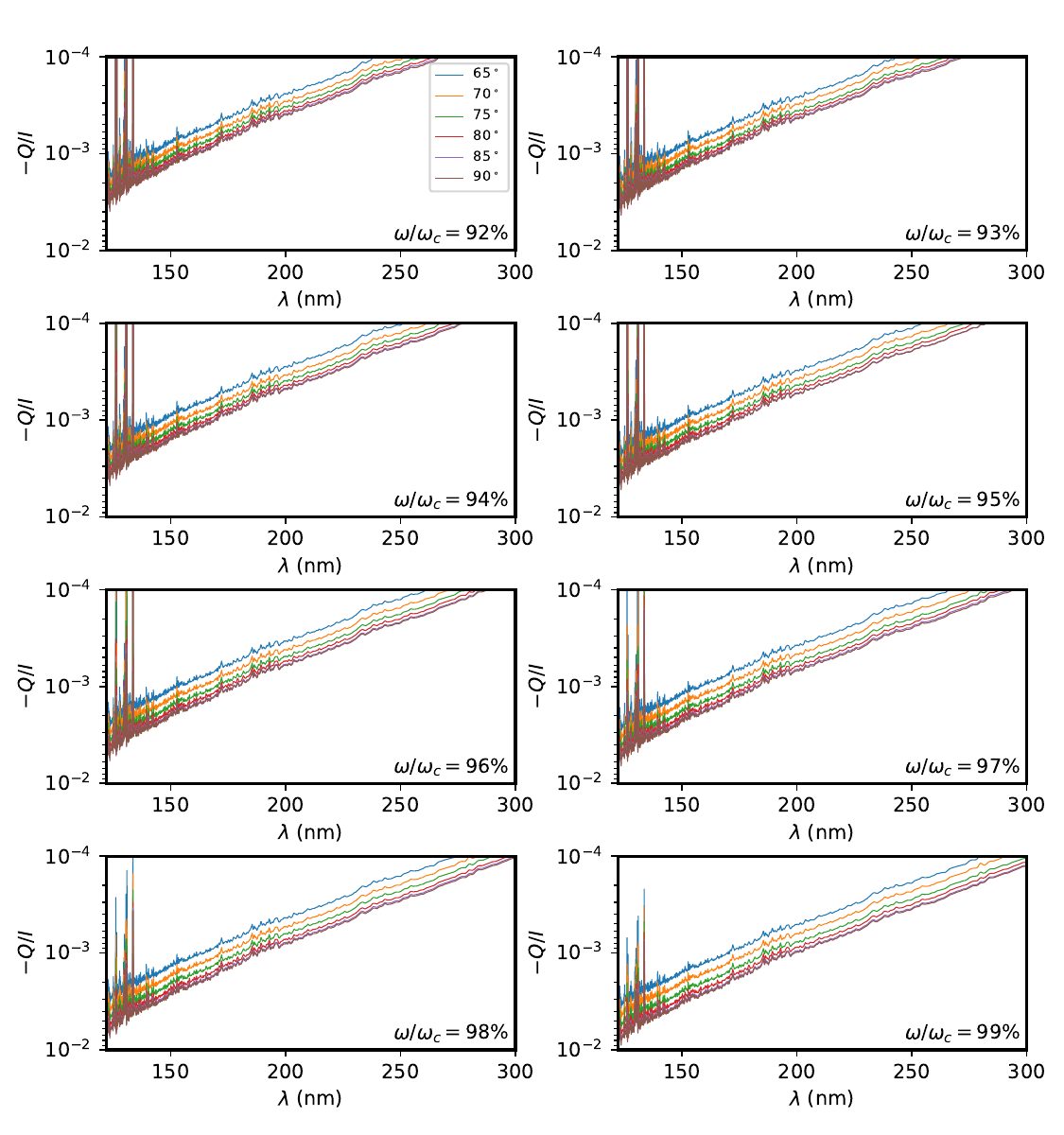}
    \caption{SYNSPEC/VLIDORT UV $Q/I$ polarization spectra for 48 models smoothed and binned to the resolution of Polstar's Channel 2. The models are those for Regulus derived from photometry and spectroscopy, to be compared to broadband optical polarization measurements of the star in \citet{Cotton2017}. Each panel represents a different rotational speed ($\omega/\omega_c$), with spectra for six different inclinations per panel represented by different coloured lines -- the key is given in the upper left panel. The spectra are on a log scale with larger negative polarization at the bottom.}
    \label{fig:Regulus_Q_Grid}
\end{figure*}

In Figure~\ref{fig:Regulus_Q_Grid} we present UV $Q/I$ polarization spectra corresponding to the same Regulus 2D parameter grid used in \citet[Supplementary Figure 3]{Cotton2017}.  Some polarization spectra have similar shapes and magnitudes in the polarization continua. In particular, there are only minor differences in the gradient between models corresponding to a particular inclination for a given value of $\omega/\omega_c$ and models with a higher inclination but lower $\omega/\omega_c$. Models with the same $\omega/\omega_c$ are also more difficult to distinguish at higher inclinations. However, the higher $R$ in the FUV -- where the (negative) polarization is larger --  reveals more differences in the line spectra, which reduces the SNR demands.

\begin{figure*}[!tp]
    \centering
    \includegraphics[clip, trim={0cm, 0cm, 0cm, 0cm}, width=16cm]{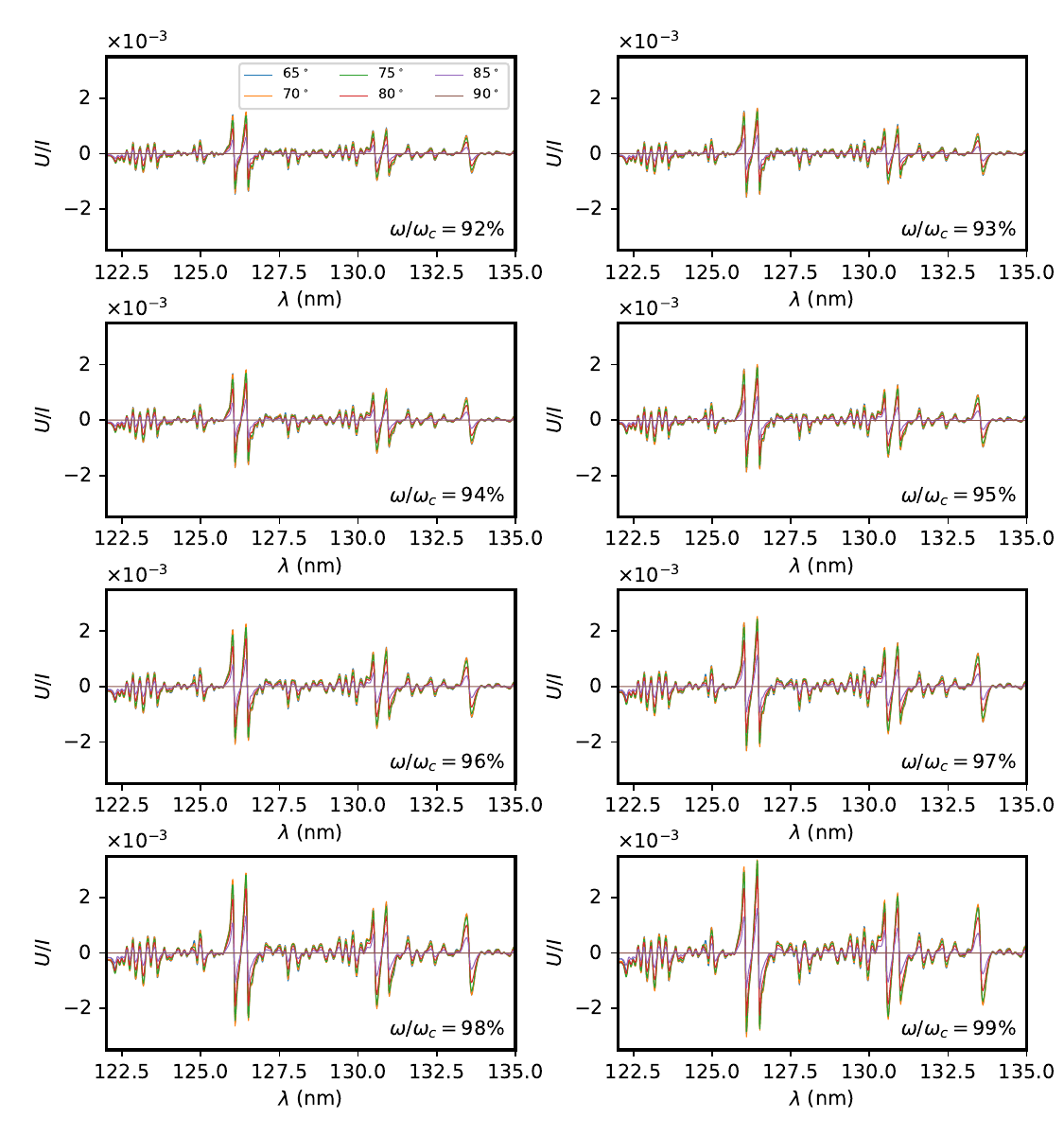}
    \caption{SYNSPEC/VLIDORT UV $U/I$ polarization spectra for 48 models smoothed and binned to the resolution of Polstar's Channel 2. The models are those for Regulus derived from photometry and spectroscopy, to be compared to broadband optical polarization measurements of the star in \citet{Cotton2017}. Each panel represents a different rotational speed ($\omega/\omega_c$), with spectra for six different inclinations per panel represented by different coloured lines -- the key is given in the upper left panel. The wavelength range has been selected to show clearly the large \"{O}hman Effect up to $\sim$4$\times$10$^{-3}$ in the FUV.}
    \label{fig:Regulus_U_Grid}
\end{figure*}

In the broadband, the polarization induced in $U/I$ from rapid rotation is essentially zero. However, it has long been predicted that line polarization will be seen in both $Q/I$ and $U/I$ as a result of considering the rotational Doppler shift where the system is inclined \citep{Ohman1946}. The \textit{\"{O}hman Effect} is only substantial at very short UV wavelengths \citep{Collins1991b}, but its impact on the $U/I$ spectrum offers a way to reduce model degeneracy at high inclinations, as shown in Figure \ref{fig:Regulus_U_Grid}.

\begin{figure}[!htp]
    \includegraphics[clip, trim={0.4cm, 0.8cm, 0.1cm, 2.1cm}, width=8cm]{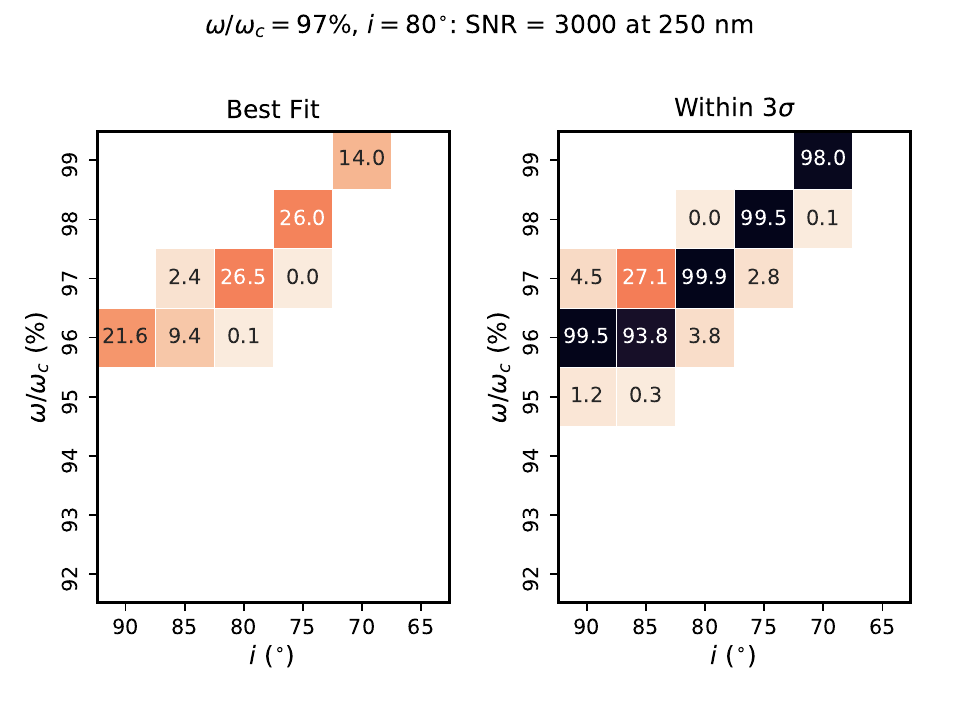}
    \includegraphics[clip, trim={0.4cm, 0.8cm, 0.1cm, 2.1cm}, width=8cm]{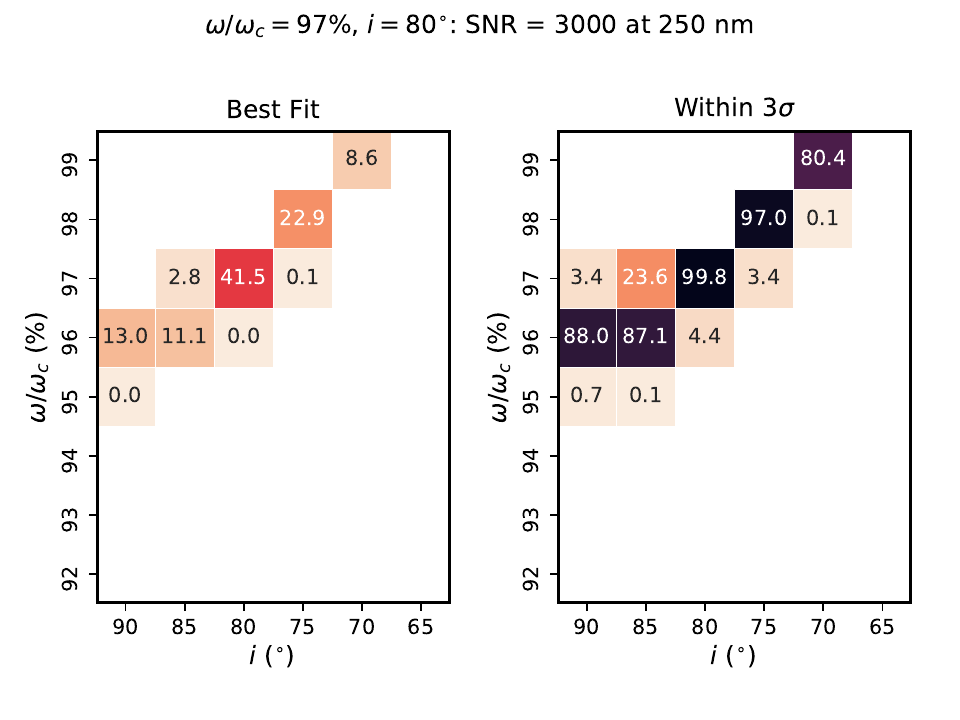}
    \includegraphics[clip, trim={0.4cm, 0.8cm, 0.1cm, 2.1cm}, width=8cm]{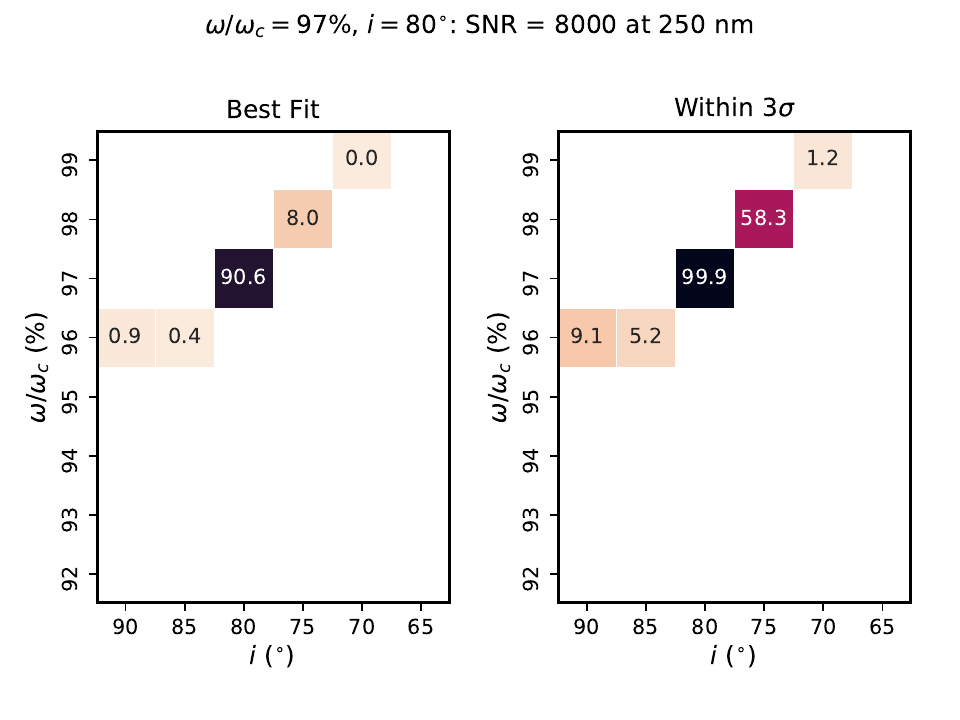}
    \caption{Tests of Polstar's sensitivity to rotation induced polarization. 10,000 sets of simulated data were generated, by adding Gaussian noise to a `Regulus' model with $\omega/\omega_c=97\%$ and $i=80^\circ$. Each data set was then compared to the grid of 48 models shown and the $\chi^2$-statistic calculated. Coloured boxes indicate a model that was either a best fit (left) or within 3$\sigma$ (right) of the best fit model for any of the 10,000 trials; the numbers in the boxes are the percentage probability of the model being the best fit (left) or within 3$\sigma$ (right) for any given trial. The top and middle rows correspond to a simulated SNR of 3,000 at 250~nm, and the lower row a SNR of 8,000 at 250~nm. In the upper row is the result of comparing only $Q/I$. Improved precision is achieved by comparing both $Q/I$ and $U/I$ as was done for the middle and lower rows.}
    \label{fig:Regulus_Expt}
\end{figure}

To examine Polstar's ability to distinguish between models of different inclination and $\omega/\omega_c$ we carried out a brief experiment. We assumed $\omega/\omega_c=97\%$ and $i=80^{\circ}$ (the closest grid model to the parameters determined in \citealp{Cotton2017}) and then generated 10,000 sets of test data by adding Gaussian noise appropriate for photon shot noise for set SNR values. SNRs of 3,000 and 8,000 at 250~nm were chosen to correspond to 10~min and 60~min exposures of the faintest Bn stars in the sample, as presented in Figure~\ref{fig:brightness_hist}. For each simulated data set the $\chi^2$-statistic for each model in the grid was calculated. Figure \ref{fig:Regulus_Expt} shows the results in terms of how often (as a percentage) each model produced the best fit to the simulated data (left) and how often each model was within 3-$\sigma$ ($\chi^2\le9$) of the best-fit model (right).

The top row of Figure~\ref{fig:Regulus_Expt} corresponds to calculating $\chi^2$ for $Q/I$ only. The middle row corresponds to the same SNR (3,000), but both $Q/I$ and $U/I$ are used to calculate $\chi^2$. It can be seen that this offers an improvement in recovering the modelled parameters, and that a large majority of the time the inclination can be determined within $\sim10^\circ$ and $\omega/\omega_c$ to within  $\sim$2$\%$. By increasing the SNR to 8,000 (as in the bottom row of Figure~\ref{fig:Regulus_Expt}) this is improved to 5$^\circ$ and 1$\%$ respectively. Parameters leading to smaller polarizations -- lower inclinations, slower rotating, later spectral types -- will be less precisely determined for the same SNR. On the other hand, more luminous objects will produce higher polarizations, and larger SNRs will be achieved for most of the Polstar Bn and Be sample. Hence, we anticipate being able to meaningfully determine the rotation rate for the majority of these stars with reasonable exposure times.

A fuller assessment of precision should also consider interstellar polarization, but in the UV it is not as much of a concern. If the level of interstellar polarization is small or comparable to the intrinsic polarization, it can be simultaneously solved for, potentially with a small loss of precision in the determination of stellar parameters \citep{Bailey2020, Lewis2019}. However, supporting ground-based optical multi-band polarization measurements facilitate independent determination of the level of interstellar polarization. Standard techniques like the use of field star polarization can be used to help characterize and remove the interstellar polarization component (see \textit{e.g.,} \citealt{Wisniewski2003,Wisniewski2006,Wisniewski2007}).  Typically, in the optical the intrinsic photospheric polarization component will be small in comparison to the interstellar polarization for stars at a significant distance  ($\gtrsim$100\,pc); it can therefore often be neglected to determine the interstellar polarization of diskless stars. The fine details of interstellar polarization in the UV is a matter of ongoing investigation including for Polstar, for example, see this volume, ``Ultraviolet Spectropolarimetry with Polstar: Interstellar Medium Science'', \citet{AnderssonTC}. However, interstellar polarization is in general well approximated by the Serkowski Law \citep{Serkowski1975}, which has only two or three parameters \citep{Wilking1980, Whittet1992} -- the wavelength of maximum polarization, $\lambda_{max}$, the maximum polarization, $p_{max}$, and sometimes a `constant' $K$, where $\lambda_{max}$ is overwhelmingly observed to be at visual wavelengths \citep{Wilking1982}. Once these parameters are solved for, the Serkowski curve can be extrapolated into the UV accurately.

In practice, good measurements for magnitude across a wide wavelength range, and in the UV in particular, are desirable for precision in setting up the parameter grid \citep{Cotton2017, Bailey2020}. The UV spectral photometry that will be obtained by Polstar are therefore an asset for this purpose, as well as for improving estimates of $v\sin{i}$. Thus, with only the addition of readily available \textit{Gaia} parallaxes, Polstar becomes the ideal platform to facilitate measurements of rotation in the Bn stars in particular. For the Be stars, the often large polarization produced by the disk is a complicating factor. However, a portion of the Be stars will be diskless at the time of observation. For stars in emission (\textit{i.e.} having a disk), the two polarization components can be disentangled, since as the UV wavelength dependence of polarization predicted for rapid rotation from the photosphere is different to that produced by scattering from the disk \citep{Bjorkman1991}.

\subsection{Model the wind/disk properties of the sample stars}
\label{disks}

The spectropolarimetric observations obtained by Polstar will probe the circumstellar environment of Be/Bn stars, revealing details of their (polar) wind, disk geometry, and wind/disk interactions. Through detailed modeling of the resulting data, we will be able to place strong constraints on disk base density and the corresponding temperature profile. Modeling can also determine properties of star-disk interaction, mediated by radiation and/or stellar wind (from the Be star and/or any companions). Monte Carlo radiative transfer (MCRT) codes have been used extensively to study the disks of Be stars \citep{Wood1996b,Wood1996a,Wood1997,Hoffman2003,Carciofi2006,Carciofi2008}. In addition, recent MCRT simulations have investigated the optical and IR polarization produced by the wind structures around massive, mass-losing stars \citep{Shrestha2018,Shrestha2021}. MCRT methods are necessary for these scenarios because the high optical depths that often exist in Be disks give rise to multiple-scattering effects that modify the polarization beyond typical analytical results \citep{Wood1996a,Shrestha2018}. The Polstar team includes several MCRT experts who will extend these existing simulation techniques into the UV for comparison with Polstar data.

\begin{figure}[t]
    \centering
    \includegraphics[scale=0.4]{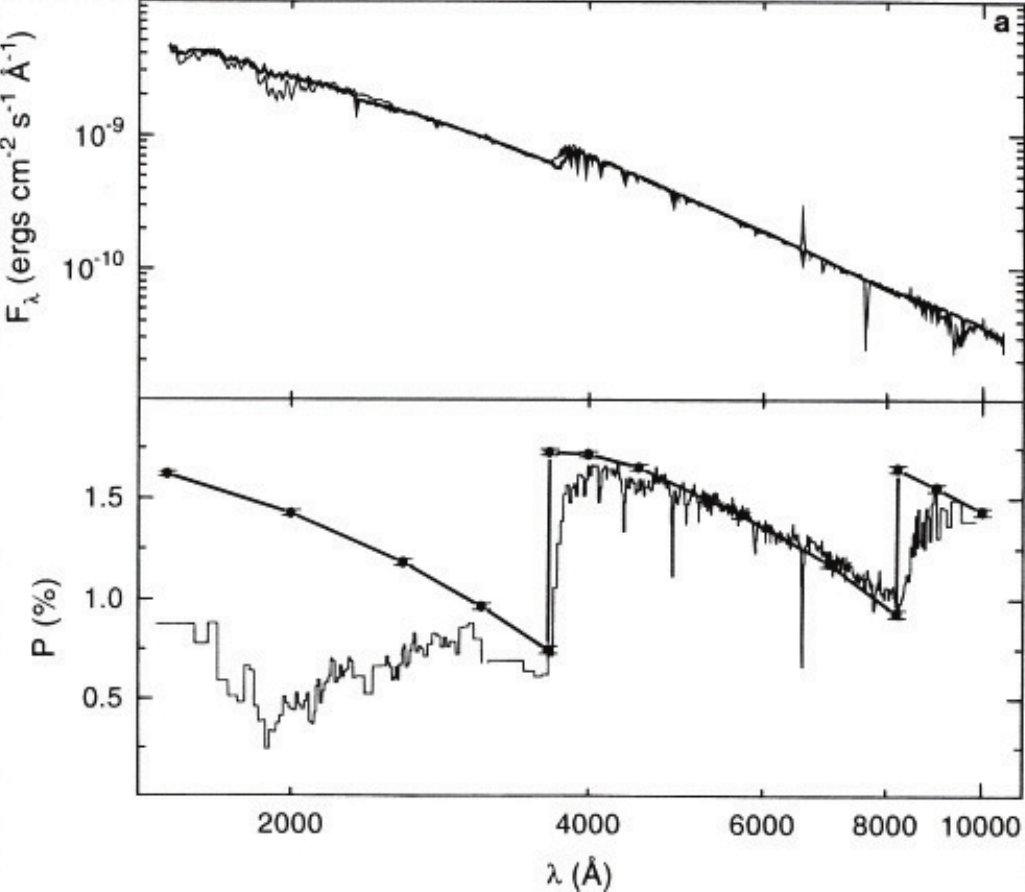}
    \caption{Thin disk model fit (thick line) to the observations (thin line) of the UV through optical spectropolarimetry of $\zeta$ Tau. The data have been corrected for interstellar reddening and polarization. Reproduced with permission from \citet{Wood1997}, figure 4, panel a.}
    \label{fig:zetatau}
\end{figure}

\citet{Wood1997} demonstrated the utility of a full-spectrum polarimetric model in analyzing the disk geometry of $\zeta$ Tau. See Figure~\ref{fig:zetatau}. Comparing the sizes of the polarization Balmer and Paschen jumps and the various continuum slopes in their model with polarimetric data from UV to IR allowed these authors to determine that the $\zeta$ Tau disk is geometrically thin 
and is consistent with either a Keplerian or wind-compressed disk but more recent work has ruled out the latter.  Also, note in Figure~\ref{fig:zetatau}
at wavelengths short-ward of the Balmer jump in the UV spectral range, that the model predictions (thick line) are over-estimated compared to the observations (thin line) due to the presence of many UV metal lines. This spectral region will provide diagnostics for the disk and stellar surface. The high-quality UV polarized spectra Polstar will obtain will enable more detailed constraints on the geometry of disks and other circumstellar configurations (\textit{e.g.}, bipolar outflows; \citealt{Schulte-Ladbeck1992}), particularly when combined with sophisticated MCRT models taking into account non-LTE effects \citep{Carciofi2006,Carciofi2008}, line blanketing, and geometries more complex than simple 2D disks \citep{Shrestha2018,Shrestha2021}.

\begin{figure*}[t]
    \centering
    \includegraphics[scale = 0.3]{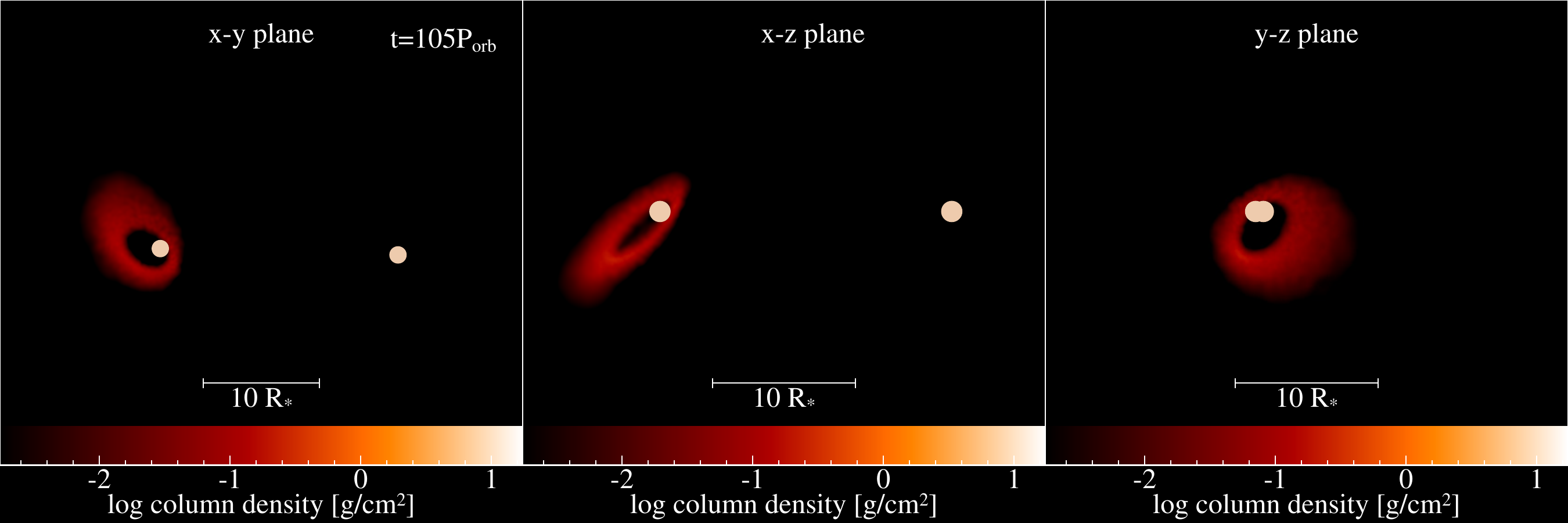}
    \caption{A 30 day binary period, $\rm 60^o$ simulation after 100 $P_{orb}$ of disk building and 5 $P_{orb}$ of disk dissipation. Left to right shows the $x-y$, $x-z$, and $y-z$ planes. The equal mass primary and secondary stars are represented by white circles, and the disc is coloured by its column density, indicated by the colour bars under each window. The scale bar in each window indicates the length of 10 primary stellar radii ($R_*$). Figure reproduced from \citet{Suffak2021}, figure 8, panel 3.}
    \label{fig:30/60_105}
\end{figure*}

Obtaining spectropolarimetric data in the UV range will complement existing optical and IR polarimetric observations and allow us to probe the stellar surface and innermost parts of the disk where material is ejected and re-accreted. For example, Figure~\ref{fig:30/60_105} shows an eccentric hole that has developed in a misaligned Be binary system. The gravitational influence of the companion causes the particles in the disk to follow eccentric orbits resulting in re-accretion onto the star when the disk particles collide with the star \citep{Suffak2021}. By constraining the disk geometry, Polstar will be able to detect the effects of the binary companion on the disk such as these holes predicted in misaligned systems, and shown in shown in Figure~\ref{fig:30/60_105}. In fact, these features only appear in binary systems where the influence of the companion causes gas to orbit on eccentric paths which, in some cases, impact the stellar surface. For isolated Be stars, these cavities in the disk are not predicted. It is well documented that Be star disks empty from the inside, however, in single stars systems, as the gas accretes, gas moves inward from greater radial distance to fill the void, such that a true gap never develops. Interestingly, in misaligned binary systems, this phenomenon is predicted to occur even when the disk is building.

\begin{figure*}[t]
    \centering
    \includegraphics[scale = 0.23]{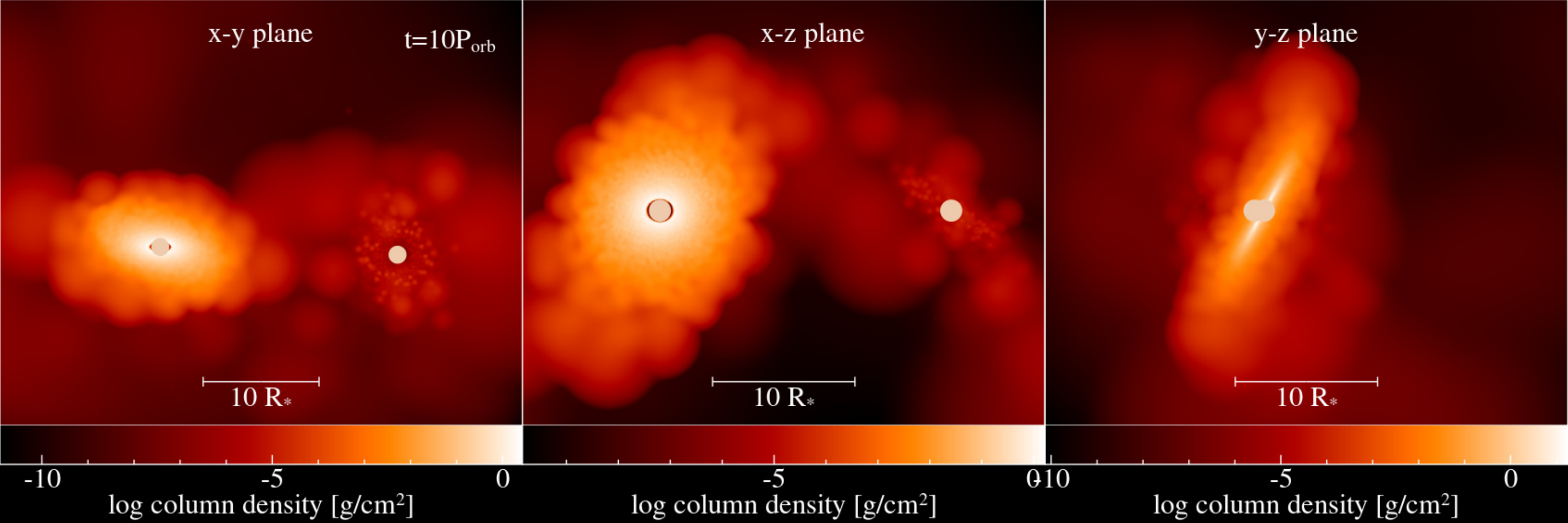}
    \caption{Snapshots of the x-y, x-z, and y-z planes after 10 orbital periods of an equal-mass binary simulation with a circular 30 day orbital period and 60 degree misalignment angle, including {\bf a} particle splitting feature. The stars are denoted by white circles, and the disk is coloured according to its column density, indicated by the colour bars below each panel. The scale bar in each window indicates the length of 10 primary stellar radii.}
    \label{secondarydisk}
\end{figure*}

Figure~\ref{secondarydisk} shows a simulation of a binary system where after a short time a disk has developed around the companion star from ejected material from the Be star. This means that if the material from an evolved companion initially caused the spin-up of the Be star, then the Be star can potentially donate a small amount of this material back (\textit{i.e.}, it is not lost from the system). 
This Figure also demonstrates that there may be opportunities to probe the interaction of the polar wind with the circumstellar material. The UV resonance lines will provide details about changes in mass outflow between visits to our targets to be investigated. Also, the UV lines Polstar obtains will potentially allow the interaction regions between the disk and winds to be mapped.

\section{Conclusions}
\label{conclude}

UV spectropolarimetry, and in particular the Polstar mission, is uniquely suited to addressing fundamental questions regarding the rapidly rotating OB population. In Section~\ref{results}, we demonstrate how stellar and binary properties of the Be/Bn stars can be revealed. Specifically, Section~\ref{Be+sd0} shows how high-resolution high-SNR UV spectroscopy can be used to probe the known Be+sdO population. The discovery of a sample of companion sdO stars will further constrain the multiplicity of the Be/Bn populations and the importance of past binary interactions for these stars. Also, their properties including orbital details, facilitated by multiple observations, can be determined. Knowing their RV's will allow the spectra to be properly shifted leading to a greater precision of determined stellar parameters. Finally, inclinations can be found using polarization to probe disk scattering. For the double-lined systems, knowing the Be/Bn and the sdO properties will place strong constraints on evolutionary pathways as well.

Polstar's low-resolution UV spectropolarimetry serves as a precise probe of the surface gradients and asymmetries that arise from rapid rotation, and is an excellent observable from which to determine rotation rates.  Having a large sample of Be/Bn stars with modeled disk and rotational properties (Sections~\ref{sdO_models},~\ref{rotation}, and~\ref{disks}), we can then answer questions as to the geometry of their orbits. With RV solutions for both components of double-lined binaries, we can then use the derived orbital/disk inclinations to determine the primary and secondary star masses.

With a distribution of stellar masses for both component stars, we will use models for binary populations to test the formation histories of the Be and Bn stars, such as the BPASS grids calculated by \citet{Eldridge2017} and \citet{Stanway2018}, which are accessible through a Python interface \citep{Stevance2020}. These types of comparisons are useful in that they provide the means to interpret the multiplicity fractions of the Be/Bn stars. This will reveal the importance of the binary verus single-star evolutionary channels. These comparisons are only accessible through the high-quality spectropolarimetry of Polstar, which enables the determination of inclinations of these systems and promotes the discovery of the relatively faint companion sdO stars that seem increasingly common based on the results of \citet{Wang2018}. 

Finally, we demonstrate in Section~\ref{disks} how the circumstellar environment can be used to investigate the polar wind, disk geometry, and disk wind interactions. Given that polarization probes the innermost disk (which, arguably, is the least well understood region), we will be able to put strong constraints on its density.  Our expertise with MCRT methods will be beneficial because the high optical depths that often exist in Be disks give rise to multiple-scattering effects that modify the polarization beyond typical analytical results.

In summary, the potential observations provided by the Polstar mission will address key outstanding issues related to the Be/Bn stellar population and the evolution of massive stars, in general.

\bibliography{main}
\pagebreak
\newpage
\section*{Statements \& Declarations}

The authors would like to thank the anonymous referee for a thorough reading, suggestions and detailed comments that helped to improve this paper.

This research has made use of NASA's Astrophysics Data System and the SIMBAD database, operated at CDS, Strasbourg, France.
This work utilized the BeSS database, operated at LESIA, Observatoire de Meudon, France: \url{http://basebe.obspm.fr} and Astropy, \url{http://www.astropy.org} a community-developed core Python package for Astronomy \citep{astropy2013, astropy2018}.

\subsection{Funding}
CEJ wishes to acknowledge support through the Natural Sciences and Engineering Research Council of Canada, NSERC. JLH acknowledges support from the National Science Foundation under award AST-1816944 and from the University of Denver via a 2021 PROF award. JL-B acknowledges support from FAPESP (grant 2017/23731-1). 
DVC wishes to thank Prof. Jeremy Bailey for his assistance in setting up the SYNSPEC/VLIDORT program and Prof. Ian Howarth for useful discussions.
Y.N. acknowledges support from the Fonds National de la Recherche Scientifique (Belgium), the European Space Agency (ESA) and the Belgian Federal Science Policy Office (BELSPO) in the framework of the PRODEX Programme (contracts linked to XMM-Newton and Gaia).
GJP gratefully acknowledges support from NASA grant 80NSSC18K0919 and STScI grants HST-GO-15659.002 and HST-GO-15869.001. DJH acknowledges support from STScI grant HST-AR-16131.001-A.
ACC acknowledges support from CNPq (grant 311446/2019-1) and FAPESP (grants 2018/04055-8 and 2019/13354-1). RI acknowledges funding support from a grant by the National Science Foundation (NSF), AST-2009412.

\subsection{Competing Interests}
The authors have no relevant financial or non-financial interests to disclose.

\subsection{Author Contributions} All authors shared ideas to motivate our work and to concretely show that UV spectroscopy and spectropolarimetry is key to understanding many aspects of massive stars, especially the Be and Bn stars and their companions.  All authors contributed to the writing of the paper and commented on the manuscript. JLB analyzed grids of OB+sdO star synthetic spectra (computed by DJH) to determine the detectability of faint sdO companions, and curated the target list. 
DVC completed the polarized radiative transfer modelling of rapid rotation.

\subsection{Data Availability}
The simulated data generated for Section \ref{sec:rot_calc} is available upon reasonable request. Data sharing is otherwise not applicable to this article as no other datasets were generated or analysed during the current study.

\clearpage
\newpage

\onecolumn
\centering
\appendix

\section*{Affiliations}

$^{1}${\orgdiv{Department of Physics and Astronomy, Western University, London, ON N6A 3K7, Canada}} 

$^{2}${\orgdiv{Homer L. Dodge Department of Physics and Astronomy, University of Oklahoma, 440 W. Brooks Street, Norman, OK 73019, USA}}

\noindent
$^{3}${\orgdiv{Monterey Institute for Research in Astronomy, 200 Eighth Street, Marina, CA, 93933, USA}}


\noindent
$^{4}${\orgdiv{Western Sydney University, Locked Bag 1797, Penrith-South DC, NSW 1797, Australia}}

\noindent
$^{5}${\orgdiv{University of Southern Queensland, Centre for Astrophysics, Toowoomba, QLD 4350, Australia}}

\noindent
$^{6}${\orgdiv{FNRS/Universit\'e de Li\`ege, All\'ee du 6 Ao\^ut 19c (B5c), B-4000 Sart Tilman, Li\`ege, Belgium}}

\noindent
$^{7}${\orgdiv{Department of Physics and Astronomy, University of Southern California, Los Angeles, CA 90089, USA}}

\noindent
$^{8}${\orgdiv{Department of Physics and Astronomy \& Pittsburgh Particle Physics, Astrophysics and Cosmology Center (PITT PACC), University of Pittsburgh, 3941 O’Hara Street, Pittsburgh, PA 15260, USA}}

\noindent
$^{9}${\orgdiv{LESIA, Paris Observatory, PSL University, CNRS, Sorbonne University, Universit\'e Paris Cit\'e, 5 place Jules Janssen, 92195 Meudon, France}}

\noindent
$^{10}${\orgdiv{Department of Physics and Astronomy, Embry-Riddle Aeronautical University, 3700 Willow Creek Rd, Prescott, AZ, 86301, USA}}

\noindent
$^{11}${\orgdiv{Department of Physics and Astronomy, University of Denver, 2112 E. Wesley Ave., Denver, CO 80208, USA}}

\noindent
$^{12}${\orgdiv{Instituto de Astronomia, Geof{\' i}sica e Ci{\^e}ncias Atmosf{\'e}ricas, Universidade de S{\~ a}o Paulo, Rua do Mat{\~ a}o 1226, Cidade Universit{\' a}ria, 05508-900 S{\~a}o Paulo, SP, Brazil}}

\noindent

\noindent
$^{13}${\orgdiv{Department of Physics and Astronomy, University of Iowa, 203 Van Allen Hall, Iowa City, IA, 52242}}

\noindent
$^{14}${\orgdiv{Department of Physics and Astronomy, East Tennessee State University,Johnson City, TN 37615, USA}}

\noindent
$^{15}${\orgdiv{GSFC, NASA}}

\clearpage
\section{Target List} \label{sec:TL}

\vspace{5mm}

\topcaption{ Target list. The listed UV flux values (at 1500 \AA\, and 2500 \AA\,) are in units of erg/s/cm$^{2}$/\AA $\times 10^{-10}$, and are estimated from IUE data whenever possible. If not observed by IUE, these flux values are interpolated based on the V$_{mag}$ and the spectral type. The `Class' column describes the type of system as follows. `sdO' systems are confirmed Be+sdO binaries where the sdO stars has been directly detected. `gCas' are the X-ray emitting $\gamma$ Cas analogs. All `sdO' and `gCas' systems are also Be stars. The table is sorted by class, and then by V$_{mag}$ within each class. Spectral types are from the literature. The `ch. 1 $t_{exp}$' is the exposure time, in seconds, for one full Polstar observation (made up of six sub-exposures) and the corresponding SNR. The `ch. 2 $t_{exp}$' is the exposure time needed in channel 2 to deliver the polarization precision given in the next column (`ch. 2 prec.'). }

{\footnotesize
\begin{supertabular}{ccccccccccc}

\hline
\multicolumn{1}{ c }{ ID } & \multicolumn{1}{c}{alt ID} & \multicolumn{1}{c}{V$_{\rm mag}$}  & \multicolumn{1}{c}{Flux at   } & \multicolumn{1}{c}{Flux at   } & \multicolumn{1}{c}{ch. 1      } & \multicolumn{1}{c}{ch. 1   }  & \multicolumn{1}{c}{ch. 2      } & \multicolumn{1}{c}{ch. 2             } & \multicolumn{1}{c}{ST} & \multicolumn{1}{c}{Class}  \\
\multicolumn{1}{ c }{    } & \multicolumn{1}{c}{      } & \multicolumn{1}{c}{             }  & \multicolumn{1}{c}{1500 \AA  } & \multicolumn{1}{c}{ 2500 \AA } & \multicolumn{1}{c}{ $t_{exp}$ } & \multicolumn{1}{c}{ SNR    }  & \multicolumn{1}{c}{ $t_{exp}$ } & \multicolumn{1}{c}{prec.             } & \multicolumn{1}{c}{  } & \multicolumn{1}{c}{     }  \\
\multicolumn{1}{ c }{    } & \multicolumn{1}{c}{      } & \multicolumn{1}{c}{             }  & \multicolumn{2}{c}{(erg/s/cm$^{2}$/\AA $\times 10^{-10}$)}        & \multicolumn{1}{c}{   (s)   } & \multicolumn{1}{c}{        }  & \multicolumn{1}{c}{    (s)    } & \multicolumn{1}{c}{($\times 10^{-3}$)} & \multicolumn{1}{c}{  } & \multicolumn{1}{c}{     }  \\
\hline\\

\tablehead{
\hline
\multicolumn{1}{ c }{ ID } & \multicolumn{1}{c}{alt ID} & \multicolumn{1}{c}{V$_{\rm mag}$}  & \multicolumn{1}{c}{Flux at   } & \multicolumn{1}{c}{Flux at   } & \multicolumn{1}{c}{ch. 1      } & \multicolumn{1}{c}{ch. 1   }  & \multicolumn{1}{c}{ch. 2      } & \multicolumn{1}{c}{ch. 2             } & \multicolumn{1}{c}{ST} & \multicolumn{1}{c}{Class}  \\
\multicolumn{1}{ c }{    } & \multicolumn{1}{c}{      } & \multicolumn{1}{c}{             }  & \multicolumn{1}{c}{1500 \AA  } & \multicolumn{1}{c}{ 2500 \AA } & \multicolumn{1}{c}{ $t_{exp}$ } & \multicolumn{1}{c}{ SNR    }  & \multicolumn{1}{c}{ $t_{exp}$ } & \multicolumn{1}{c}{prec.             } & \multicolumn{1}{c}{  } & \multicolumn{1}{c}{     }  \\
\multicolumn{1}{ c }{    } & \multicolumn{1}{c}{      } & \multicolumn{1}{c}{             }  & \multicolumn{2}{c}{(erg/s/cm$^{2}$/\AA $\times 10^{-10}$)}        & \multicolumn{1}{c}{   (s)   } & \multicolumn{1}{c}{        }  & \multicolumn{1}{c}{    (s)    } & \multicolumn{1}{c}{($\times 10^{-3}$)} & \multicolumn{1}{c}{  } & \multicolumn{1}{c}{     }  \\
\hline\\
}

HD 10516 & phi Per & 4.06 & 15.0 & 5.5 & 190 & 200  & 1498  & 0.20  & B1.5 V:e-shell & sdO \\ 
HD 200120 & 59 Cyg & 4.75 & 11.0 & 3.1 & 260 & 200  & 2658  & 0.20  & B1.5Vnne & sdO \\ 
HD 41335 & HR 2142 & 5.21 & 4.0 & 2.0 & 410 & 150  & 3600  & 0.21  & B1.5IV-Vnne & sdO \\ 
HD 157042 & iota Ara & 5.25 & 4.0 & 2.0 & 410 & 150  & 3600  & 0.21  & B2Vnne & sdO \\ 
HD 200310 & 60 Cyg & 5.43 & 5.0 & 1.7 & 330 & 150  & 3600  & 0.23  & B1Ve & sdO \\ 
HD 137387 & kap01 Aps & 5.50 & 4.0 & 1.6 & 410 & 150  & 3600  & 0.24  & B2Vnpe & sdO \\ 
HD 58978 & FY CMa & 5.56 & 6.0 & 1.8 & 490 & 200  & 3600  & 0.23  & B0.5IVe & sdO \\ 
HD 60855 & V378 Pup & 5.70 & 2.0 & 0.9 & 830 & 150  & 3600  & 0.32  & B2Ve & sdO \\ 
HD 43544 & HR 2249 & 5.90 & 3.0 & 1.1 & 550 & 150  & 3600  & 0.29  & B2/B3Ve & sdO \\ 
HD 194335 & V2119 Cyg & 5.90 & 3.0 & 1.1 & 550 & 150  & 3600  & 0.29  & B2IIIe & sdO \\ 
HD 113120 & HD 113120 & 6.00 & 1.6 & 0.7 & 1040 & 150  & 3600  & 0.36  & B2IVne & sdO \\ 
HD 152478 & V846 Ara & 6.30 & 0.9 & 0.4 & 820 & 100  & 3600  & 0.51  & B3Vnpe & sdO \\ 
HD 51354 & QY Gem & 7.20 & 0.6 & 0.3 & 1230 & 100  & 3600  & 0.52  & B3ne & sdO \\ 
HD 29441 & V1150 Tau & 7.60 & 0.3 & 0.1 & 2460 & 100  & 3600  & 0.87  & B2.5Vne & sdO \\ 
HD 55606 &  & 9.04 & 0.1 & 0.1 & 1540 & 50  & 3600  & 1.24  & B0.5Vnnpe & sdO \\ 
\hline 
HD 5394 & gamma Cas & 2.39 & 107.0 & 39.0 & 60 & 300  & 211  & 0.20  & B0.5 Ivpe & gCas \\ 
HD 212571 & pi Aqr & 4.60 & 10.0 & 3.1 & 660 & 300  & 2658  & 0.20  & B1 III-Ive & gCas \\ 
HD 110432 & BZ Cru & 5.31 & 1.2 & 0.7 & 1887 & 175  & 3600  & 0.36  & B0.5IVpe & gCas \\ 
HD 44458 & FR CMa & 5.55 & 2.0 & 0.9 & 1133 & 175  & 3600  & 0.32  & B1.5IVe & gCas \\ 
HD 120991 & V767 Cen & 6.10 & 1.2 & 0.8 & 1887 & 175  & 3600  & 0.34  & B2Ve & gCas \\ 
HD 45995 &  & 6.14 & 1.5 & 0.9 & 1507 & 175  & 3600  & 0.32  & B1.5Vne & gCas \\ 
HD 183362 &  & 6.34 & 1.1 & 0.5 & 2060 & 175  & 3600  & 0.43  & B3Ve & gCas \\ 
HD 45314 &  & 6.64 & 0.6 & 0.2 & 1927 & 125  & 3600  & 0.60  & O9:npe & gCas \\ 
HD 157832 & V750 Ara & 6.66 & 0.6 & 0.4 & 1927 & 125  & 3600  & 0.51  & B2ne & gCas \\ 
HD 12882 &  & 7.62 & 0.2 & 0.2 & 7200 & 139  & 3600  & 0.63  & B2.5III:[n]e+ & gCas \\ 
HD 119682 &  & 7.90 & 0.2 & 0.2 & 7200 & 139  & 3600  & 0.73  & B0Ve & gCas \\ 
HD 220058 &  & 8.59 & 0.1 & 0.1 & 7200 & 99  & 3600  & 1.01  & B2 & gCas \\ 
BD+43 3913 &  & 8.91 & 0.2 & 0.1 & 7200 & 135  & 3600  & 1.18  & B1.5V:nnep & gCas \\ 
HD 161103 &  & 9.13 & 0.2 & 0.1 & 7200 & 122  & 3600  & 1.30  & B0.5III/IVe & gCas \\ 
HD 162718 &  & 9.16 & 0.0 & 0.0 & 7200 & 68  & 3600  & 2.25  & B3/5ne & gCas \\ 
BD+47 3129 &  & 9.27 & 0.1 & 0.0 & 7200 & 115  & 3600  & 1.38  & B0 & gCas \\ 
HD 316568 &  & 9.66 & 0.1 & 0.0 & 7200 & 96  & 3600  & 1.59  & B3 & gCas \\ 
HD 90563 &  & 9.86 & 0.1 & 0.0 & 7200 & 87  & 3600  & 1.75  & B2Ve & gCas \\ 
HD 130437 &  & 10.04 & 0.1 & 0.0 & 7200 & 80  & 3600  & 30.49  & B1Ve & gCas \\ 
\hline 
HD 10144 & Achernar & 0.46 & 26.0 & 12.0 & 253 & 300  & 687  & 0.20  & B6Vpe & Be \\ 
HD 127972 & Eta Cen & 2.31 & 92.0 & 35.0 & 67 & 300  & 235  & 0.20  & B2 Ve & Be \\ 
HD 143275 & del Sco & 2.32 & 70.0 & 30.0 & 93 & 300  & 274  & 0.20  & B0IV & Be \\ 
HD 105435 & del Cen & 2.52 & 55.0 & 23.0 & 120 & 300  & 358  & 0.20  & B2Vne & Be \\ 
HD 23630 & Eta Tau & 2.87 & 15.7 & 6.7 & 420 & 300  & 1224  & 0.20  & B7III & Be \\ 
HD 58715 & beta CMi & 2.89 & 11.0 & 5.8 & 600 & 300  & 1421  & 0.20  & B8Ve & Be \\ 
HD 37202 & Zeta Tau & 3.03 & 30.2 & 15.0 & 220 & 300  & 549  & 0.20  & B1 IVe\_Shell & Be \\ 
HD 205021 & Bet Cep & 3.23 & 35.3 & 12.9 & 187 & 300  & 636  & 0.20  & B0.5IIIs & Be \\ 
HD 89080 & ome Car & 3.33 & 5.5 & 3.5 & 1207 & 300  & 2357  & 0.20  & B8IIIe & Be \\ 
HD 11415 & eps Cas & 3.37 & 31.0 & 11.4 & 213 & 300  & 724  & 0.20  & B3Vp\_sh & Be \\ 
HD 120324 & Mu Cen  & 3.43 & 33.5 & 10.8 & 193 & 300  & 765  & 0.20  & B2 Vnpe & Be \\ 
HD 217675 & Omi And & 3.62 & 7.9 & 3.4 & 840 & 300  & 2445  & 0.20  & B6IV/V\_sh & Be \\ 
HD 23302 & 17 Tau & 3.70 & 7.3 & 3.1 & 907 & 300  & 2632  & 0.20  & B6IIIe & Be \\ 
HD 56139 & 28 CMa & 3.82 & 13.0 & 7.5 & 507 & 300  & 1096  & 0.20  & B2.5Ve & Be \\ 
HD 166014 & Omi Her & 3.83 & 2.4 & 1.9 & 927 & 175  & 3600  & 0.22  & B9.5III & Be \\ 
HD 50013 & Kappa CMa & 3.89 & 28.9 & 10.0 & 227 & 300  & 824  & 0.20  & B1.5 Ve & Be \\ 
HD 109387 & kap Dra & 3.89 & 8.0 & 3.5 & 827 & 300  & 2357  & 0.20  & B6IIIe & Be \\ 
HD 135734 & mu Lup & 4.00 & 2.1 & 1.7 & 1087 & 175  & 3600  & 0.24  & B8Ve & Be \\ 
HD 25940 & 48 Per & 4.03 & 16.9 & 6.2 & 393 & 300  & 1330  & 0.20  & B3Ve & Be \\ 
HD 45542 & nu Gem & 4.14 & 6.0 & 2.6 & 1107 & 300  & 3173  & 0.20  & B6IVe & Be \\ 
HD 23480 & 23 Tau & 4.18 & 4.7 & 2.0 & 980 & 250  & 3600  & 0.21  & B6IV(e) & Be \\ 
HD 149630 & Sig Her & 4.20 & 1.7 & 1.4 & 1300 & 175  & 3600  & 0.26  & B9V & Be \\ 
HD 173948 & lam Pav & 4.21 & 14.4 & 5.3 & 460 & 300  & 1567  & 0.20  & B2Ve & Be \\ 
HD 22192 & psi Per & 4.23 & 5.5 & 2.2 & 1207 & 300  & 3600  & 0.20  & B5Ve & Be \\ 
HD 6811 & Phi And & 4.25 & 4.4 & 1.9 & 1047 & 250  & 3600  & 0.22  & B5IIIe & Be \\ 
HD 33328 & Lam Eri & 4.27 & 13.6 & 5.0 & 487 & 300  & 1659  & 0.20  & B2III(e)p & Be \\ 
HD 102776 & j Cen & 4.31 & 13.1 & 4.8 & 507 & 300  & 1722  & 0.20  & B3Ve & Be \\ 
HR 5704 & gam Cir & 4.35 & 4.0 & 1.7 & 1147 & 250  & 3600  & 0.23  & B5IVe & Be \\ 
HD 202904 & ups Cyg & 4.42 & 11.8 & 4.3 & 560 & 300  & 1906  & 0.20  & B2Vne & Be \\ 
HD 148184 & Chi Oph & 4.43 & 11.7 & 4.3 & 567 & 300  & 1924  & 0.20  & B2Vne & Be \\ 
HD 63462 & Omi Pup & 4.49 & 12.0 & 4.1 & 553 & 300  & 2032  & 0.20  & B1IVe & Be \\ 
HD 75311 & f Car & 4.49 & 11.1 & 4.1 & 600 & 300  & 2032  & 0.20  & B3Vne & Be \\ 
HD 4180 & omi Cas & 4.50 & 3.5 & 1.5 & 1320 & 250  & 3600  & 0.25  & B5IIIe & Be \\ 
HD 198183 & lam Cyg & 4.54 & 3.4 & 1.4 & 1367 & 250  & 3600  & 0.25  & B5Ve & Be \\ 
HD 205637 & Eps Cap & 4.55 & 10.5 & 3.8 & 633 & 300  & 2147  & 0.20  & B3V & Be \\ 
HD 37490 & Ome Ori & 4.59 & 10.1 & 3.7 & 660 & 300  & 2230  & 0.20  & B3Ve & Be \\ 
HD 164284 & 66 Oph & 4.60 & 7.0 & 2.3 & 947 & 300  & 3588  & 0.20  & B2Ve & Be \\ 
HD 112078 & lam Cru & 4.60 & 10.0 & 3.7 & 667 & 300  & 2252  & 0.20  & B3Vne & Be \\ 
HD 78764 & E Car & 4.65 & 6.0 & 3.5 & 1107 & 300  & 2357  & 0.20  & B2(IV)n & Be \\ 
HD 56014 & 27 Cma & 4.65 & 3.0 & 1.3 & 1513 & 250  & 3600  & 0.26  & B4Ve\_sh & Be \\ 
HD 57150 & ups01 Pup & 4.67 & 9.4 & 3.4 & 707 & 300  & 2398  & 0.20  & B2V+B3IVne & Be \\ 
HD 209409 & omi Aqr & 4.69 & 2.7 & 1.4 & 833 & 175  & 3600  & 0.26  & B7IVe & Be \\ 
HD 192685 & QR Vul & 4.75 & 8.7 & 3.2 & 760 & 300  & 2583  & 0.20  & B3Ve & Be \\ 
HD 83953 & I Hya & 4.76 & 2.8 & 1.2 & 820 & 175  & 3600  & 0.28  & B5V & Be \\ 
HD 68980 & r Pup & 4.77 & 8.6 & 3.1 & 773 & 300  & 2632  & 0.20  & B1Ve & Be \\ 
HD 158643 & 51 Oph & 4.81 & 1.0 & 0.8 & 1160 & 125  & 3600  & 0.34  & A0V & Be \\ 
HD 92938 & V518 Car & 4.82 & 8.2 & 3.0 & 813 & 300  & 2754  & 0.20  & B3Ve & Be \\ 
HD 20336 & BK Cam & 4.85 & 7.9 & 2.9 & 833 & 300  & 2833  & 0.20  & B2.5Vne & Be \\ 
HD 142983 & 48 Lib & 4.87 & 3.5 & 1.0 & 1320 & 250  & 3600  & 0.30  & B5IIIp\_sh & Be \\ 
HD 191610 & 28 Cyg & 4.93 & 4.5 & 1.8 & 1027 & 250  & 3600  & 0.23  & B2.5Ve & Be \\ 
HD 35439 & 25 Ori & 4.96 & 7.0 & 2.6 & 947 & 300  & 3137  & 0.20  & B1Vne & Be \\ 
HD 187811 & 12 Vul & 4.96 & 7.2 & 2.6 & 927 & 300  & 3137  & 0.20  & B2.5Ve & Be \\ 
HD 212076 & 31 Peg & 4.99 & 6.0 & 2.0 & 1107 & 300  & 3600  & 0.21  & B2IV-Ve & Be \\ 
HD 24479 & HD 24479 & 5.04 & 0.8 & 0.6 & 1440 & 125  & 3600  & 0.38  & B9IV & Be \\ 
HD 124367 & V795 Cen & 5.07 & 2.1 & 0.9 & 1093 & 175  & 3600  & 0.32  & B4Vne & Be \\ 
HD 32343 & 11 Cam & 5.08 & 6.4 & 2.4 & 1033 & 300  & 3496  & 0.20  & B3Ve & Be \\ 
HD 23862 & Pleione & 5.10 & 1.2 & 0.6 & 1880 & 175  & 3600  & 0.39  & B8Vne & Be \\ 
HD 131492 & tet Cir & 5.11 & 6.3 & 2.3 & 1060 & 300  & 3597  & 0.20  & B2IIIe & Be \\ 
HD 48917 & 10 Cma & 5.17 & 5.9 & 2.2 & 1120 & 300  & 3600  & 0.21  & B2Ve & Be \\ 
HD 71510 & HR 3330 & 5.17 & 5.9 & 2.2 & 1120 & 300  & 3600  & 0.21  & B3IVe & Be \\ 
HD 203467 & 6 Cep & 5.18 & 5.9 & 2.1 & 1133 & 300  & 3600  & 0.21  & B3IVe & Be \\ 
HD 189687 & 25 Cyg & 5.19 & 5.8 & 2.1 & 1147 & 300  & 3600  & 0.21  & B3IVe & Be \\ 
HD 169985 & d Ser & 5.32 & 0.1 & 0.1 & 1233 & 50  & 3600  & 0.92  & A0Vs+G:III & Be \\ 
HD 167128 & HR 6819 & 5.36 & 5.0 & 1.8 & 927 & 250  & 3600  & 0.22  & B2Ve & Be \\ 
HD 142184 & HD 142184 & 5.40 & 4.8 & 1.8 & 967 & 250  & 3600  & 0.23  & B2V & Be \\ 
HD 28497 & 228 Eri & 5.41 & 5.5 & 2.0 & 1207 & 300  & 3600  & 0.21  & B2(V)ne & Be \\ 
HD 180968 & ES Vul & 5.43 & 4.7 & 1.7 & 993 & 250  & 3600  & 0.23  & B0.5IV & Be \\ 
HD 217050 & EW Lac & 5.43 & 2.6 & 0.7 & 867 & 175  & 3600  & 0.36  & B4IIIpe & Be \\ 
HD 58155 & NO Cma & 5.43 & 4.7 & 1.7 & 993 & 250  & 3600  & 0.23  & B3Ve & Be \\ 
HD 105521 & V817 Cen & 5.50 & 4.4 & 1.6 & 1053 & 250  & 3600  & 0.24  & B3IVe & Be \\ 
HD 91120 & HD 91120 & 5.58 & 0.5 & 0.4 & 1520 & 100  & 3600  & 0.49  & B8/9IV/V & Be \\ 
HD 144 & 10 Cas & 5.59 & 0.5 & 0.4 & 1520 & 100  & 3600  & 0.49  & B9IIIe & Be \\ 
HD 46860 & mu Pic & 5.64 & 3.8 & 1.4 & 1200 & 250  & 3600  & 0.26  & B9IVne & Be \\ 
HD 49131 & HP Cma & 5.68 & 3.7 & 1.4 & 1247 & 250  & 3600  & 0.26  & B1.5Vne & Be \\ 
HD 214168 & 8 Lac A & 5.69 & 4.0 & 1.3 & 1153 & 250  & 3600  & 0.26  & B1Vne & Be \\ 
HD 36576 & 120 Tau & 5.69 & 1.5 & 0.8 & 1500 & 175  & 3600  & 0.34  & B2IV-Ve & Be \\ 
HD 23016 & 13 Tau & 5.69 & 1.2 & 0.5 & 1920 & 175  & 3600  & 0.43  & B7Ve & Be \\ 
HD 169033 & HD 169033 & 5.70 & 1.2 & 0.5 & 1940 & 175  & 3600  & 0.43  & B5V & Be \\ 
HD 88661 & HR 4009 & 5.75 & 2.2 & 1.1 & 1027 & 175  & 3600  & 0.29  & B5Vne & Be \\ 
HD 142926 & 4 Her & 5.75 & 0.4 & 0.3 & 1760 & 100  & 3600  & 0.53  &  B9pe & Be \\ 
HD 72067 & HR 3356 & 5.81 & 3.3 & 1.2 & 1407 & 250  & 3600  & 0.28  & B2/3Ve & Be \\ 
HD 30076 & 56 Eri & 5.81 & 1.6 & 1.2 & 1400 & 175  & 3600  & 0.28  & B2(V)nne & Be \\ 
HD 66194 & V374 Car & 5.81 & 3.3 & 1.2 & 1407 & 250  & 3600  & 0.28  & B3Vne & Be \\ 
HD 54309 & FV Cma & 5.83 & 3.2 & 1.2 & 1433 & 250  & 3600  & 0.28  & B3Vne & Be \\ 
HD 208682 & HD 208682 & 5.86 & 3.1 & 1.1 & 1473 & 250  & 3600  & 0.28  & B2Ve & Be \\ 
HD 63215 & V392 Pup & 5.87 & 1.0 & 0.4 & 1160 & 125  & 3600  & 0.46  & B5Ve & Be \\ 
HD 129954 & CO Cir & 5.88 & 3.1 & 1.1 & 1500 & 250  & 3600  & 0.28  & B2Ve & Be \\ 
HD 149671 & eta01 TrA & 5.88 & 1.0 & 0.4 & 1180 & 125  & 3600  & 0.47  & B7IVe & Be \\ 
HD 174237 & CX Dra & 5.90 & 3.0 & 1.1 & 1527 & 250  & 3600  & 0.29  & B3+F5III & Be \\ 
HD 32991 & 105 Tau & 5.92 & 3.0 & 1.1 & 760 & 175  & 3600  & 0.29  & B2Ve & Be \\ 
HD 158220 & V862 Ara & 5.98 & 0.9 & 0.4 & 1280 & 125  & 3600  & 0.49  & B7IIIe & Be \\ 
HD 185037 & 11 Cyg & 6.03 & 0.3 & 0.3 & 2300 & 100  & 3600  & 0.60  & B8Vne & Be \\ 
HD 29866 & HD 29866 & 6.08 & 0.3 & 0.2 & 2400 & 100  & 3600  & 0.62  & B8V & Be \\ 
HD 183656 & V923 Aql & 6.08 & 0.8 & 0.3 & 1420 & 125  & 3600  & 0.51  & B7III & Be \\ 
HD 23552 & HD 23552 & 6.14 & 0.3 & 0.2 & 2540 & 100  & 3600  & 0.63  & B8V & Be \\ 
HD 183537 & 7 Vul & 6.33 & 0.6 & 0.3 & 1780 & 125  & 3600  & 0.58  & B5Vne & Be \\ 
HD 58050 & OT Gem & 6.41 & 1.9 & 0.7 & 1200 & 175  & 3600  & 0.36  & B2Ve & Be \\ 
HD 65875 & HD 65875 & 6.59 & 1.6 & 0.6 & 1400 & 175  & 3600  & 0.39  & B2.5Ve & Be \\ 
HD 170235 & HD 170235 & 6.59 & 1.6 & 0.6 & 1400 & 175  & 3600  & 0.39  & B1Vnne & Be \\ 
HD 24534 & X Per & 6.72 & 0.2 & 0.1 & 1067 & 50  & 3600  & 0.83  & O9.5III & Be \\ 
HD 6226 & V442 And & 6.82 & 0.3 & 0.5 & 2460 & 100  & 3600  & 0.44  & B2.5III & Be \\ 
HD 203699 & HD 203699 & 6.86 & 1.2 & 0.5 & 1800 & 175  & 3600  & 0.45  & B2.5IVne & Be \\ 
HD 162732 & 88 Her & 6.89 & 0.4 & 0.2 & 1900 & 100  & 3600  & 0.74  & B6IIInp\_sh & Be \\ 
HD 218393 & KX And & 6.92 & 1.2 & 0.4 & 1920 & 175  & 3600  & 0.46  & B3pe+K1III & Be \\ 
HD 11606 & V777 Cas & 7.02 & 1.1 & 0.4 & 2100 & 175  & 3600  & 0.48  & B2Vne & Be \\ 
HD 17520 & HD 17520 & 8.24 & 0.4 & 0.1 & 2100 & 100  & 3600  & 0.85  & O8V+O9:Ve & Be \\ 
\hline 
HD 87901 & alf Leo & 1.40 & 50.0 & 25.0 & 120 & 300  & 329  & 0.20  & B8IVn & Bn \\ 
HD 169022 & eps Sgr & 1.85 & 15.1 & 12.0 & 440 & 300  & 688  & 0.20  & B9IVp & Bn \\ 
HD 135742 & bet Lib & 2.62 & 15.0 & 8.0 & 440 & 300  & 1030  & 0.20  & B8Vn & Bn \\ 
HD 106490 & del Cru & 2.75 & 54.9 & 20.1 & 120 & 300  & 410  & 0.20  & B2IVn & Bn \\ 
HD 177724 & zet Aql & 2.99 & 5.3 & 4.2 & 1240 & 300  & 1968  & 0.20  & A0IV-Vnn & Bn \\ 
HD 136298 & del Lup & 3.19 & 36.6 & 13.4 & 180 & 300  & 614  & 0.20  & B1.5IVn & Bn \\ 
HD 143118 & eta Lup & 3.41 & 29.9 & 11.0 & 220 & 300  & 751  & 0.20  & B2.5IVn & Bn \\ 
HD 177756 & lam Aql & 3.43 & 3.5 & 2.8 & 1300 & 250  & 2949  & 0.20  & B8.5V & Bn \\ 
HD 125238 & iot Lup & 3.53 & 26.8 & 9.8 & 240 & 300  & 839  & 0.20  & B2.5IVn & Bn \\ 
HD 17573 & 41 Ari & 3.59 & 3.0 & 2.4 & 1520 & 250  & 3435  & 0.20  & B8Vn & Bn \\ 
HD 158094 & del Ara & 3.62 & 7.0 & 3.5 & 940 & 300  & 2357  & 0.20  & B8Vn & Bn \\ 
HD 98718 & pi Cen & 3.64 & 7.7 & 3.3 & 860 & 300  & 2489  & 0.20  & B5Vn & Bn \\ 
HD 21364 &  & 3.75 & 2.6 & 2.1 & 860 & 175  & 3600  & 0.21  & B9Vn & Bn \\ 
HD 213998 & eta Aqr & 4.03 & 2.0 & 1.6 & 1100 & 175  & 3600  & 0.24  & B8/9V & Bn \\ 
HD 128345 & rho Lup & 4.05 & 16.6 & 6.1 & 400 & 300  & 1355  & 0.20  & B3/4Vn & Bn \\ 
HD 144294 & tet Lup & 4.20 & 12.0 & 5.3 & 540 & 300  & 1558  & 0.20  & B2.5Vn & Bn \\ 
HD 164577 & 68 Oph & 4.43 & 1.4 & 1.1 & 1600 & 175  & 3600  & 0.29  & A0.5Van & Bn \\ 
HD 33802 & iot Lep & 4.45 & 1.4 & 1.1 & 1640 & 175  & 3600  & 0.29  & B7.5Vn & Bn \\ 
HD 64503 & b Pup & 4.47 & 11.2 & 4.1 & 580 & 300  & 2003  & 0.20  & B2Vn & Bn \\ 
HD 225132 & 2 Cet & 4.56 & 1.3 & 0.8 & 1740 & 175  & 3600  & 0.35  & B9IVn & Bn \\ 
HD 142114 & 2 Sco  & 4.59 & 6.0 & 3.7 & 1100 & 300  & 2230  & 0.20  & B2.5Vn & Bn \\ 
HD 141637 & 1 Sco & 4.63 & 7.0 & 2.1 & 940 & 300  & 3600  & 0.21  & B1.5Vn & Bn \\ 
HD 113703 & f Cen & 4.69 & 2.9 & 1.3 & 760 & 175  & 3600  & 0.27  & B4Vn & Bn \\ 
HD 24072 & HR 1190 & 4.72 & 1.1 & 0.9 & 2100 & 175  & 3600  & 0.33  & B9.5Van & Bn \\ 
HD 176638 & zet CrA & 4.72 & 1.0 & 0.8 & 2260 & 175  & 3600  & 0.33  & B9.5Vann & Bn \\ 
HD 93194 & HD 4205 & 4.79 & 5.0 & 2.0 & 1320 & 300  & 3600  & 0.21  & B3/5Vn & Bn \\ 
HD 108257 & G Cen & 4.81 & 8.3 & 3.0 & 800 & 300  & 2726  & 0.20  & B3Vn & Bn \\ 
HD 93607 & HR 4222 & 4.85 & 2.5 & 1.1 & 880 & 175  & 3600  & 0.29  & B4Vn & Bn \\ 
HD 15130 &  & 4.87 & 0.9 & 0.7 & 1220 & 125  & 3600  & 0.35  & A0V & Bn \\ 
HD 32309 &  & 4.88 & 0.9 & 0.7 & 1240 & 125  & 3600  & 0.35  & B9V & Bn \\ 
HD 184915 & kap Aql & 4.96 & 4.0 & 1.8 & 1140 & 250  & 3600  & 0.23  & B0.5IIIn & Bn \\ 
HD 42167 & tet Col & 4.99 & 2.2 & 1.0 & 1020 & 175  & 3600  & 0.31  & B7IIIn & Bn \\ 
HD 184606 & 9 Vul & 5.00 & 0.8 & 0.7 & 1380 & 125  & 3600  & 0.37  & B8IIIn & Bn \\ 
HD 43445 &  & 5.00 & 0.8 & 0.7 & 1380 & 125  & 3600  & 0.37  & B9V & Bn \\ 
HD 181296 & eta Tel  & 5.02 & 0.8 & 0.6 & 1400 & 125  & 3600  & 0.38  & A0V+M7/8V & Bn \\ 
HD 20418 & 31 Per & 5.03 & 2.1 & 0.9 & 1040 & 175  & 3600  & 0.32  & B5Vn & Bn \\ 
HD 196740 & 28 Vul & 5.05 & 2.1 & 0.9 & 1060 & 175  & 3600  & 0.32  & B5IVn & Bn \\ 
HD 135240 & del Cir & 5.09 & 6.4 & 2.3 & 1040 & 300  & 3532  & 0.20  & O8V & Bn \\ 
HD 172777 & lam CrA & 5.11 & 0.8 & 0.6 & 1540 & 125  & 3600  & 0.39  & A0/1V & Bn \\ 
HD 74753 & D Vel & 5.16 & 3.0 & 2.2 & 1540 & 250  & 3600  & 0.20  & B1/2II/III(n) & Bn \\ 
HD 18331 &  & 5.16 & 0.7 & 0.6 & 1600 & 125  & 3600  & 0.40  & A1V & Bn \\ 
HD 26793 &  & 5.22 & 0.6 & 0.5 & 1920 & 125  & 3600  & 0.41  & B9Vn & Bn \\ 
HD 168905 &  & 5.23 & 5.6 & 2.1 & 1180 & 300  & 3600  & 0.21  & B3V & Bn \\ 
HD 222847 & i01 Aqr & 5.24 & 0.7 & 0.5 & 1720 & 125  & 3600  & 0.42  & B9V & Bn \\ 
HD 212710 &  & 5.26 & 0.6 & 0.5 & 1920 & 125  & 3600  & 0.42  & B9.5Vn & Bn \\ 
HD 34863 &  & 5.28 & 0.6 & 0.5 & 1800 & 125  & 3600  & 0.42  & B7/8V & Bn \\ 
HD 20809 & HR 1011 & 5.30 & 2.2 & 0.7 & 1020 & 175  & 3600  & 0.36  & B5V & Bn \\ 
HD 93540 & HR 4219 & 5.34 & 1.6 & 0.7 & 1400 & 175  & 3600  & 0.36  & B6Vnn & Bn \\ 
HD 107696 &  & 5.37 & 1.6 & 0.7 & 1420 & 175  & 3600  & 0.37  & B7Vn & Bn \\ 
HD 136849 & 50 Boo & 5.37 & 0.6 & 0.5 & 1960 & 125  & 3600  & 0.44  & B9Vn & Bn \\ 
HD 208321 &  & 5.44 & 0.6 & 0.4 & 2080 & 125  & 3600  & 0.46  & A3V & Bn \\ 
HD 19134 & 52 Ari & 5.47 & 1.4 & 0.6 & 1580 & 175  & 3600  & 0.39  & B7Vn & Bn \\ 
HD 189395 &  & 5.50 & 0.5 & 0.4 & 2200 & 125  & 3600  & 0.47  & B9Vn & Bn \\ 
HD 35770 & 116 Tau & 5.51 & 0.5 & 0.4 & 2240 & 125  & 3600  & 0.47  & B9.5Vn & Bn \\ 
HD 560 & 34 Psc & 5.53 & 0.5 & 0.4 & 2300 & 125  & 3600  & 0.48  & B9Vn & Bn \\ 
HD 159358 &  & 5.54 & 0.5 & 0.4 & 2280 & 125  & 3600  & 0.48  & B9V & Bn \\ 
HD 21362 &  & 5.57 & 1.3 & 0.6 & 1720 & 175  & 3600  & 0.41  & B6Vn & Bn \\ 
HD 209833 & 23 Peg & 5.70 & 0.4 & 0.3 & 1680 & 100  & 3600  & 0.52  & B9Vn & Bn \\ 
HD 188293 & 57 Aql A & 5.71 & 1.1 & 0.5 & 1960 & 175  & 3600  & 0.43  & B7Vn & Bn \\ 
HD 150745 & HR 6215 & 5.73 & 3.5 & 1.3 & 1300 & 250  & 3600  & 0.27  & B2III/IV & Bn \\ 

\end{supertabular}
}

\twocolumn

\clearpage
\section{Detectability of faint sdO companions} \label{sec:detect_sdO}
\justifying

Further details are described here concerning the methods used to determine the channel 1 spectroscopic SNR needed to adequately explore the parameter space of Be/Bn+sdO binaries. 
It is worth keeping in mind that while in the known Be+sdO binaries the sdO star contributes about 2\% -- 10\% of the total flux in the UV, these cases were discovered primarily through analysis of low-SNR archival spectra which were not sensitive enough to detect fainter systems. In order to determine the binary fraction and properties of the Be/n population in general, while discovering and characterizing potential Be/n+sdO systems, 
observations should be sufficiently sensitive as to detect hot sub-luminous companions even when they contribute only $\sim$0.1\% (or a few times this) of the total UV flux. 
This is justified in Sec.~\ref{sec:BesdO_parameters} and further examined here.

\subsection{Methodology}

Synthetic composite spectra were used to emulate Polstar observations of Be/n+sdO binaries using the non-LTE radiative code CMFGEN \citep{Hillier1998}. For the (O)B star component, models were generated for $\log{g}$ = 4.0 and five values of $T_{\rm eff}$ (13, 18, 20, 25, and 35 kK), roughly corresponding to spectral types of B8, B6, B4, B1, and O9. These spectra were convolved with a Gaussian so that absorption lines were broadened to correspond approximately to $v\sin{i}$ = 250 km s$^{-1}$. 
Ten sdO model spectra were generated for $\log{g}$ = 4.5 and 5.0, and $T_{\rm eff}$ = 30, 35, 40, 45, and 50 kK, broadened to correspond to $v\sin{i}$ = 15 km s$^{-1}$. 
The B and sdO spectra were then co-added over a range of luminosity 
($L_{\rm sdO}$ = 0.1 -- 1,000 $L_{\odot}$) and noise levels (SNR between 10 -- 1,000) for a given B star model spectrum.
For testing the relatively luminous O star, $L_{\rm sdO}$ was extended to 10,000 $L_{\odot}$, although such high sdO luminosities are not realistic.

\begin{figure*}
\centering
\includegraphics[width=0.49\textwidth]{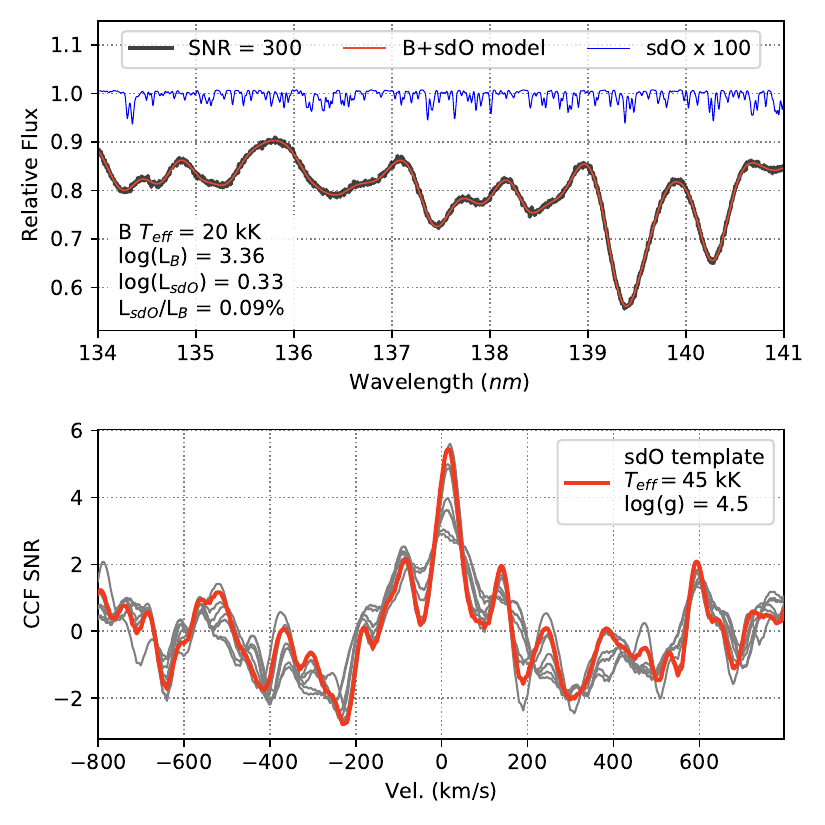}
\includegraphics[width=0.49\textwidth]{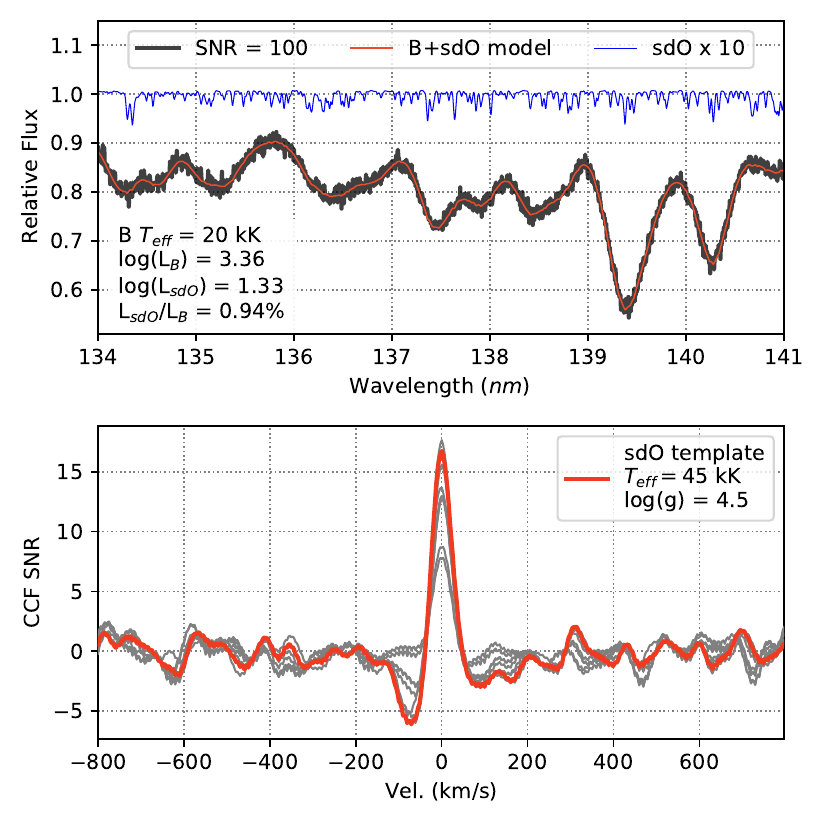}
\caption{CCF analysis for simulated Be/n+sdO spectra. The left panel illustrates a faint sdO system that is marginally detected, and in the right panel the sdO CCF signal is strong. The top panel shows the co-added model spectrum for a B+sdO system (red curve) with noise added (black curve), and also the isolated sdO spectrum (scaled by the factor in the legend). The CCF signal is shown in the lower panel for the correct sdO model (in red) and the other sdO templates (grey) with different values of  $T_{\rm eff}$ and $\log{g}$. No velocity shifts were introduced to the input spectra.   }
\label{sdO_SNR_threshold}
\end{figure*}

In nearly all trials, the features in the co-added spectra are dominated by the relatively luminous B star (see top panels of Figure~\ref{sdO_SNR_threshold}). Since the absorption lines of the rapidly-rotating B star are wide compared to the sdO lines, in effect the sdO spectrum can be isolated by convolving the co-added spectrum with a Gaussian that is wider than the sdO features but more narrow than the B star features, and then dividing the simulated data by the `smoothed' fit. This was done prior to cross-correlating the sdO template spectra. However, in practice, it may be preferable to simultaneously (or iteratively) perform a two-component CCF analysis considering both the B and sdO stars. In these tests, the input model spectrum of the B star (scaled to the correct flux ratio) could have been removed from the co-added noise-added spectrum perfectly, but such an ideal fit of the B star model to the data is not expected to be realized in practice, and simply applying a smoothing kernel is probably closer to a more realistic treatment of observational data.

A CCF analysis was then used to attempt to recover the spectroscopic signature of the sdO star. 
Figure~\ref{sdO_SNR_threshold} illustrates part of this procedure for two trials --  a threshold case and a strong detection.
 
Each sdO template was cross-correlated with the (noisy) simulated spectrum (after removing the broad features, as above). The strength of the CCF signal corresponding to the input sdO model is then used as the metric to gauge the degree of detectability of the sdO spectrum for a given combination of the adopted stellar and noise parameters.  In this application, a CCF SNR of $\gtrsim$5 can be considered a (marginal) detection, 
but a stronger signal is needed to differentiate between different sdO model spectra. 
In the majority of cases where the CCF SNR $\gtrsim$10, the templates with the correct $T_{\rm eff}$ typically have noticeably higher signals than the other models (surface gravity was harder to differentiate).
However, the details of determining the sdO stellar properties from these simulations were not investigated in detail.

\subsection{Results}

\begin{figure*}
\centering
\includegraphics[width=0.49\textwidth]{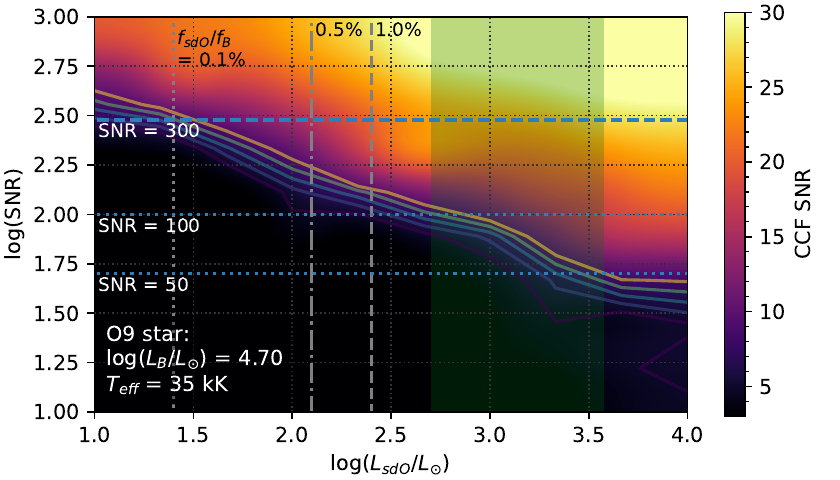}
\includegraphics[width=0.49\textwidth]{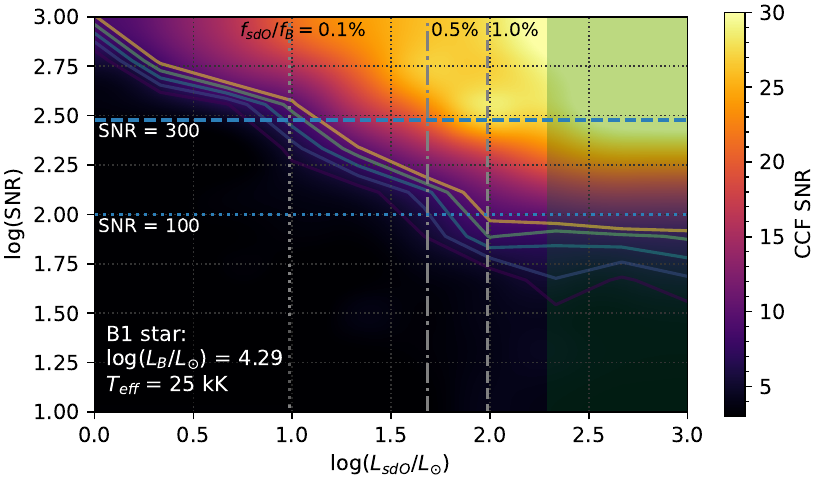}
\includegraphics[width=0.49\textwidth]{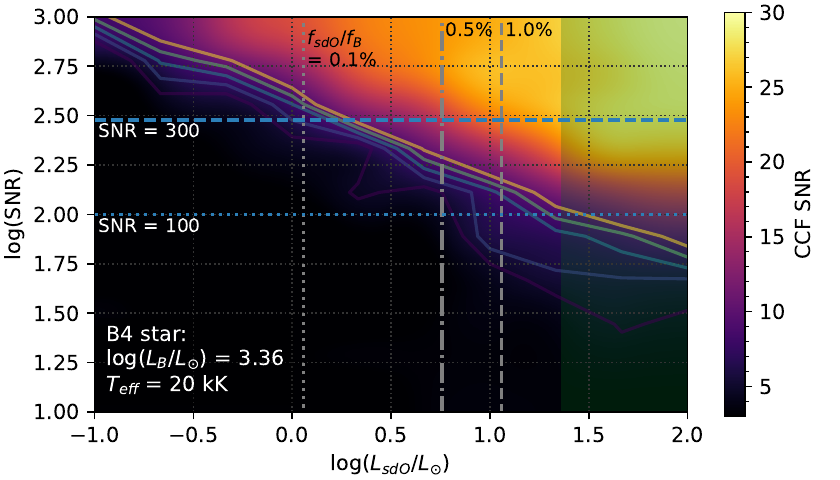}
\includegraphics[width=0.49\textwidth]{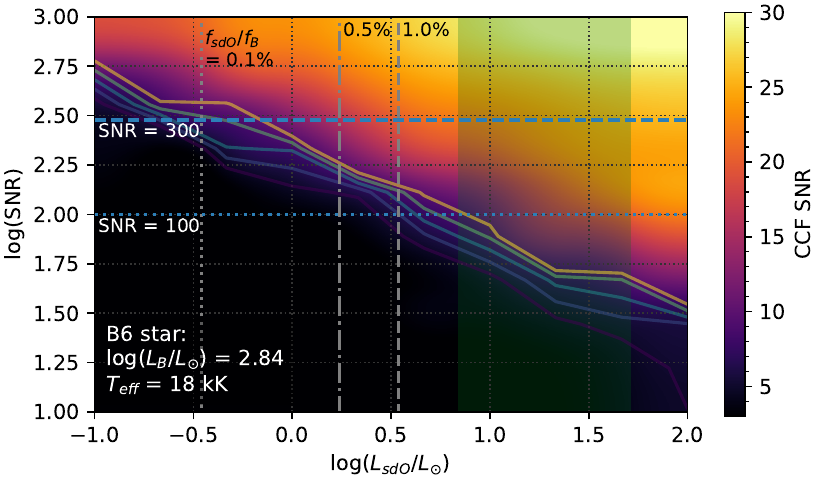}
\includegraphics[width=0.49\textwidth]{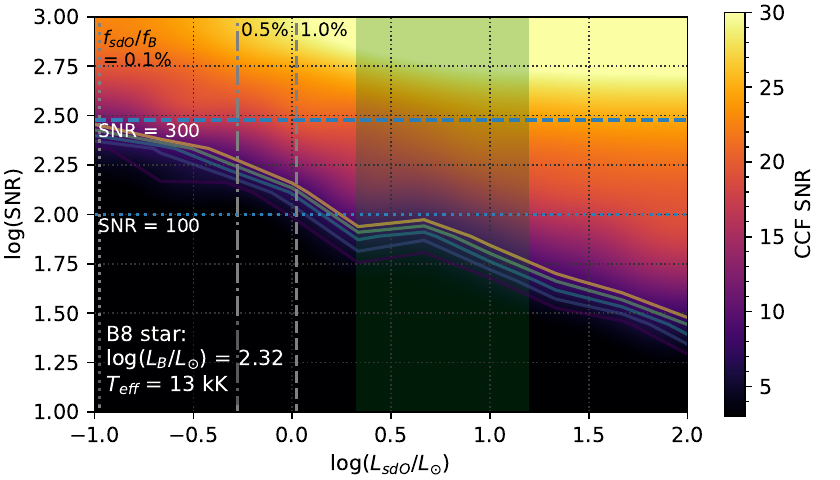}
\caption{
Simulated detection thresholds for recovering the spectroscopic signature of an sdO star in a Be/n+sdO binary. Each panel represents simulated scenarios with a rapidly-rotating B star (properties in the bottom-left corner) and an sdO star over a range of luminosity for a UV spectrum with $R = 30,000$. The contours between the dark and light regions trace the boundaries for a CCF SNR of 5 to 10, where a CCF SNR = 5 should be sufficient to detect an sdO spectrum, while higher CCF SNR values improve the ability to accurately characterize the sdO star. The green shaded rectangle marks the sdO luminosity range that corresponds to flux ratios of $f_{sdO}/f_{B} = 2 - 10\%$ (note, however, that this range corresponds to different values of $L_{\rm sdO}$ depending on the B star model used). 
The plotted range in $L_{\rm sdO}$ for each panel depends on the (O)B star luminosity, which spans over two orders of magnitude.} 
\label{sdO_space_25}
\end{figure*}

Figure~\ref{sdO_space_25} gives an overview of 
the `detectability space' in terms of UV spectroscopic SNR and sdO luminosity. For this family of tests, an sdO model with $\log{g}$ = 4.5 and $T_{\rm eff}$ = 45 kK was added to the five aforementioned (O)B star model spectra over values of $L_{\rm sdO}$ and SNR. Virtually the same detectability thresholds are found when using as input any of the other sdO spectra. 

In Figure~\ref{sdO_space_25} the lighter yellow regions (upper right in each panel) represent configurations where the sdO spectrum is detected strongly and the spectroscopic parameters can be readily determined. The dark regions in each panel represent non-detections. At a CCF SNR of $\sim$5 (in the definitions employed here) the sdO spectrum is marginally detected. At CCF SNRs of around 8 -- 10, the fidelity of the recovered spectrum becomes high enough to begin to characterize the sdO stellar properties (\textit{e.g.,} $T_{\rm eff}$ and $\log{g}$).
The contours in each panel in Figure~\ref{sdO_space_25} trace the boundaries at CCF SNR = 5, 6, 7, 8, and 9, and thus this strip and above is the region where a given combination of $L_{\rm sdO}$ and spectroscopic SNR should render a detection in a single observation.

These tests show that high SNR ($\sim$300) UV spectroscopy is sensitive enough to detect sdO stars that are $\sim$1,000 times fainter than their rapidly-rotating B-type companion. And, in the case of non-detections at this degree of precision, stringent constraints can be placed on the nature of an unseen companion (or lack thereof).
With Polstar, such observations can be obtained with reasonable exposure times for a significant fraction of the bright sample of Be/n stars listed in Appendix~\ref{sec:TL}.

\vspace{5mm}






\end{document}